\documentclass[%
 reprint,
 amsmath,amssymb,
 aps,
]{revtex4-2}

\usepackage{graphicx}% Include figure files
\usepackage{dcolumn}% Align table columns on decimal point
\usepackage{bm}% bold math
\usepackage{xcolor}
\usepackage{algorithm} 
\usepackage{algpseudocode} 
\usepackage{tabularx}

\begin{document}

%\preprint{APS/123-QED}

\title{Analysis framework for higher-order temporal correlations with applications to human heartbeats}

\author{Tibebe Birhanu}
\affiliation{Department of Physics, The Catholic University of Korea, Bucheon, Republic of Korea}

\author{Hang-Hyun Jo}
\email{h2jo@catholic.ac.kr}
\affiliation{Department of Physics, The Catholic University of Korea, Bucheon, Republic of Korea}

\date{\today}% It is always \today, today,
             %  but any date may be explicitly specified

\begin{abstract}
We propose a time series analysis framework focused on higher-order temporal correlations in the event sequence beyond the interevent time distribution by employing the burst-tree decomposition method. Bursts are clustered events that rapidly occur within shorter time periods, and they are separated by relatively longer inactive periods. The burst-tree decomposition method exactly maps the event sequence onto a tree, called a burst tree, in which each internal node represents a merge of consecutive bursts at the timescale separating those bursts. The burst tree fully reveals the hierarchical structure of bursts, hence the higher-order temporal correlations for the entire range of timescales. Those correlations are quantified using novel and existing measures derived from the burst tree, such as the burst complexity, memory coefficient for bursts, and principal and secondary cross sections of the burst-merging kernel. We apply our framework to the heartbeat time series of healthy people and of those with heart disease to reveal distinct multiscale temporal properties in physiological time series.
\end{abstract}

%\keywords{Suggested keywords}%Use showkeys class option if keyword
%display desired
\maketitle

\section{Introduction}

A complex system can be characterized by the collection of interactions between its components, which determines the system's overall behavior and dynamics~\cite{Sayama2015Introduction}. The dynamics of complex systems often shows intermittent temporal activity, which can be depicted as a bursty sequence of events. The bursty temporal pattern implies rapidly occurring events in short time periods interlaced by long inactive periods~\cite{Barabasi2005Origin, Karsai2018Bursty, Jo2023Bursty}. Examples include event sequences from solar flare activity~\cite{Wheatland1998WaitingTime, deArcangelis2006Universality} and earthquakes~\cite{Bak2002Unified, Corral2004Longterm} to neural dynamics~\cite{Beggs2003Neuronal, Petermann2009Spontaneous, Kemuriyama2010Powerlaw} and human social interactions~\cite{Barabasi2005Origin, Harder2006Correlated, Vazquez2006Modeling, Goh2008Burstiness, Rybski2009Scaling, Wu2010Evidence, Jiang2013Calling, Fournet2014Contact, Karsai2018Bursty, Jo2020Bursttree, Choi2021Individualdriven}. As these temporal patterns also show $1/f$ noise, some of such systems have been analyzed in the framework of self-organized criticality~\cite{Bak1987Selforganized, Jensen1998Selforganized, Christensen2005Complexity}. To understand the underlying mechanisms behind the complex dynamics of systems, it is of utmost importance to devise effective methods or tools characterizing temporal correlations in the data. For this, various measures and quantities have been introduced, such as the interevent time distribution, burstiness parameter, memory coefficient, burst size distribution, and autocorrelation function to name a few~\cite{Barabasi2005Origin, Goh2008Burstiness, Karsai2012Universal, Karsai2018Bursty, Jo2023Bursty}. 

The interevent time (IET) is the time interval between two consecutive events and its distribution is often found to be heavy tailed, implying the deviation from the memoryless, Poisson process~\cite{Masuda2018Gillespie}. The memory coefficient measures the correlation between two consecutive IETs~\cite{Goh2008Burstiness}, while the notion of the burst size captures the correlation between an arbitrary number of consecutive IETs~\cite{Karsai2012Universal}. That is, for a given timescale $\Delta t$, the events are clustered into multiple bursts such that the IETs between bursts are greater than $\Delta t$, and the number of events in the burst is called a burst size. Empirically found heavy-tailed distributions of burst sizes indicate the presence of higher-order temporal correlations beyond IET distributions~\cite{Karsai2012Universal, Hiraoka2025Hierarchical}.

For a more systematic approach to the analysis of empirical event sequences, the burst-tree decomposition method has been proposed~\cite{Jo2020Bursttree}. The method exactly maps the event sequence onto the tree, called a burst tree, by scanning the entire range of the timescale. Thus, the burst tree reveals the hierarchical structure of bursts across multiple timescales. In particular, each internal node of the burst tree denotes a merge of consecutive bursts as the timescale defining bursts increases. Such a merging pattern has been captured by some selection rule called a burst-merging kernel~\cite{Jo2020Bursttree, Birhanu2025Bursttree, Birhanu2025Maximum}. The burst-merging kernel dictates which bursts are merged together as the timescale increases, hence this kernel plays a crucial role in understanding the underlying mechanism of higher-order temporal correlations, often characterized by heavy-tailed burst size distributions and positive correlations between consecutive burst sizes.

In this work, we propose a time series analysis framework focused on higher-order temporal correlations beyond the IET distribution by employing the burst-tree decomposition method. Those correlations are quantified using novel and existing measures derived from the burst tree, such as the burst complexity, memory coefficient for bursts, and principal and secondary cross sections of the burst-merging kernel. For demonstration, we apply our framework to the publicly available data of heartbeat time series in health subjects exhibiting normal sinus rhythm (NSR) as well as individuals diagnosed with cardiovascular diseases such as congestive heart failure (CHF) and atrial fibrillation (AF). The burst trees derived from these data allow us to represent and analyze the complex multiscale temporal dynamics of heartbeat time series, providing a robust method for understanding the underlying physiological processes. Consequently, our study may shed light on the differences in heartbeat dynamics between healthy individuals and those with heart failure.

We brief some previous works on the heartbeat time series analysis using other methods such as the detrended fluctuation analysis (DFA), multiscale entropy (MSE), and visibility graph (VG) algorithm. These methods analyze the time series of RR intervals (RRI or interbeat intervals), i.e., the IET sequence, whereas our approach is to analyze the event sequence directly, yet the event sequence and the IET sequence are easily interchangeable. Peng et al.~\cite{Peng1994Mosaic} proposed the DFA method, which essentially quantifies the power-law scaling of the root-mean-squared fluctuation as a function of the timescale for detecting the local trend. Using the DFA method~\cite{Peng1995Quantification} it was found that heartbeat time series from the CHF group show different scaling behaviors than those from healthy people. Recently, a dynamical DFA (DDFA) was introduced to detect real-time changes in highly nonstationary RRI time series~\cite{Molkkari2020Dynamical}. Pukkila et al.~\cite{Pukkila2025Detection} adopted the DFA and DDFA to extract features from the RRI sequences. Such features were used as inputs to machine learning methods to distinguish the NSR, CHF, and AF groups. Regarding the classification or detection of subjects with heart diseases, we also find more practical machine learning methods~\cite{Ramesh2021Atrial, Cinar2021Classification}.

The MSE method based on the sample entropy~\cite{Richman2000Physiological} has been applied to the RRI time series to detect long-range temporal correlations across multiple timescales~\cite{Costa2002Multiscale, Costa2015Generalized}. It was found that the MSE curves as a function of the timescale show different behaviors depending on whether people are healthy or have heart diseases. The VG algorithm has also been applied to physiological time series~\cite{Munoz-Diosdado2023Visibility}. The VG algorithm transforms the time series to the network of nodes and links, enabling to apply various characterizations in network science to the time series analysis. All the mentioned methods have shown significant results when applied to the RRI time series, yet requiring alternative approaches to deepen the understanding of the physiological time series.

\section{Materials and Methods}\label{sec:methods}

\subsection{Data}

We analyze heartbeat time series from the datasets downloaded from the physiological data repository, PhysioNet~\cite{PhysioBank}. Every dataset has a unique identity, such as nsr2db for normal sinus rhythm (NSR), chfdb and chf2db for congestive heart failure (CHF), and afdb and ltafdb for atrial fibrillation (AF). We take three distinct groups of people for the analysis. The group of the NSR contains 54 healthy individuals from nsr2db dataset. Precisely, they are $30$ males with an age distribution ranging from $28$ to $76$ years, and $24$ females with an age distribution from $58$ to $73$ years. The CHF group consists of $44$ individuals with CHF from chfdb and chf2db datasets, of which $19$ are male and $6$ are female. The age distribution for males ranges from $22$ to $71$ years, while for females, it is from $54$ to $63$ years. The gender of the rest $19$ subjects is not known. Finally, the AF group we consider consists of $109$ individuals with AF from afdb and ltafdb datasets, in which subjects' age and sex data are not available. The duration of recording is approximately $24$ hours for the NSR, CHF and AF from ltafdb datasets, while it is around $10$ hours for the AF from the afdb dataset. Accordingly, the average number of heartbeats (or events) per individual is around $1.07\times 10^5$ for the groups of NSR, CHF, and AF from the ltafdb dataset, while it is around $4.5\times 10^4$ for the group of AF from the afdb dataset. In total, we have heartbeat time series for $207$ subjects. Prior to the analysis, each heartbeat time series was filtered to avoid the influence of outlier data. Following the previous method~\cite{Pukkila2025Detection} we discard the RR intervals (RRIs) falling outside of the range, i.e., $0.75 \eta \leq \rm RRI \leq 1.5 \eta$, where $\eta$ denotes the local median with a moving window of 31 RRIs. 

\begin{figure*}[!t]
    \centering
    \includegraphics[width=\textwidth]{fig1.pdf}
    \caption{Overview of the analysis framework for higher-order temporal correlations using the heartbeat time series of one healthy subject, referred to as ``the NSR subject 1.'' (a) A sample period of the event sequence of the NSR subject 1, where the timing of the first event was set to 0 ms. (b) The burst tree derived from the sample event sequence in (a). (c) The burst-merging kernel $K(b,b')$ estimated using the entire period of the NSR subject 1's event sequence. The white color means no data in the area. (d) The interevent time distribution $P(\tau)$ in Eq.~\eqref{eq:Ptau_define}, with the estimated values of the burstiness measure $A$ in Eq.~\eqref{eq:burstiness} and the memory coefficient $M_\tau$ in Eq.~\eqref{eq:memory}. (e) The burst size distributions $Q_{\Delta t}(b)$ for several values of $\Delta t$ after rescaling by the average burst size $\langle b\rangle_{\Delta t}$ for each $\Delta t$. (f) The burst complexity $C_{\Delta t}$ in Eq.~\eqref{eq:C_deltat} and the memory coefficient for bursts $M_{\Delta t}$ in Eq.~\eqref{eq:M_deltat} for the entire range of $\tau_{\rm min}\leq \Delta t\leq \tau_{\rm max}$. The horizontal dotted line at $M_{\Delta t}=0$ is to guide the eyes. (g) The principal and secondary diagonal cross sections $K_1(b)$ and $K_2(b)$ in Eq.~\eqref{eq:diagonals}, which are taken from the burst-merging kernel $K(b,b')$ in (c). The vertical dotted line at $b\approx 39$ in the lower panel is to guide the eyes.
    }
    \label{fig:template_fig}
\end{figure*}

\subsection{Burstiness and memory coefficient}\label{subsec:burstiness}

Let us begin by measuring the burstiness parameter and the memory coefficient of the event sequence~\cite{Goh2008Burstiness}. The heartbeat time series of each subject is given in the form of an event sequence with $n$ events, denoted as $\mathcal{E} = \{t_0, \ldots, t_{n-1}\}$, where $t_i$ represents the timing of the $i$th beat. For example, a sample period of the NSR subject 1's event sequence is shown in Fig.~\ref{fig:template_fig}(a). From the event sequence $\mathcal{E}$ we get the interevent time (IET) sequence $\{\tau_1, \ldots, \tau_{n-1}\}$ by using $\tau_i \equiv t_i - t_{i-1}$. The minimum and maximum IETs in the IET sequence are denoted by $\tau_{\rm min}$ and $\tau_{\rm max}$, respectively. The IET distribution $P(\tau)$ is then calculated as follows:
\begin{align}
    P(\tau)=\frac{1}{n-1}\sum_{i=1}^{n-1}\delta(\tau-\tau_i),
    \label{eq:Ptau_define}
\end{align}
where $\delta$ is a Dirac delta function. $P(\tau)$ for the entire period of the event sequence of the NSR subject 1 is depicted in Fig.~\ref{fig:template_fig}(d), showing one dominant peak around at 700 ms and the other smaller peak around at 1 second.

The burstiness parameter~\cite{Goh2008Burstiness} was proposed to quantify the degree of the variation of the IETs normalized by their mean, which however suffers from the finite-size effect due to the finite number of events. Later Kim and Jo introduced an alternative definition of the burstiness measure by correcting such finite-size effect~\cite{Kim2016Measuring}: 
\begin{align}
    A\equiv \frac{\sqrt{n + 1}r - \sqrt{n - 1}}{(\sqrt{n + 1}  - 2)r + \sqrt{n - 1}},
    \label{eq:burstiness}
\end{align}
where $r\equiv \sigma/\mu$ is the coefficient of variation with $\sigma$ and $\mu$ denoting the standard deviation and the mean of IETs, respectively. For a regular event sequence, $A= -1$, while the fully random event sequence leads to $A=0$. If the events are extremely bursty, one has $A\approx 1$. Next, the memory coefficient $M_{\tau}$ is defined as the Pearson correlation coefficient (PCC) between two consecutive IETs~\cite{Goh2008Burstiness}:
\begin{align}
  M_{\tau}\equiv \frac{1}{n - 2} \sum_{i = 1}^{n - 2} \dfrac{(\tau_{i} - \mu_{1})(\tau_{i+1} - \mu_{2})}{\sigma_{1}\sigma_{2}},
   \label{eq:memory}
\end{align}
where $\mu_{1}$ ($\mu_{2}$) and $\sigma_{1}$ ($\sigma_{2}$) are the mean and standard deviation of $\tau$s without $\tau_{n-1}$ ($\tau_1$) in the IET sequence, respectively. $M_\tau$ has a value in the range of $[-1,1]$; the positive $M_\tau$ implies that long (short) IETs tend to follow long (short) IETs, while it is negative for the opposite tendency. For the uncorrelated IETs, $M_\tau\approx 0$.

The estimated values of $A\approx -0.64$ and $M_\tau\approx 0.98$ for the entire period of the NSR subject 1's event sequence are shown in Fig.~\ref{fig:template_fig}(d). Although $A$ and $M_\tau$ are already informative in understanding temporal correlations, they are not enough to study the higher-order temporal correlations in the time series, requiring more systematic approaches.

\subsection{Burst-tree decomposition method}\label{subsec:decompose}

To fully characterize the higher-order temporal correlations beyond the IET distribution, the burst-tree decomposition method has been proposed~\cite{Jo2020Bursttree}, which can map the event sequence into a burst tree. This method reveals the hierarchical structure of bursts of events for a wide range of timescale $\Delta t$. For a given $\Delta t$, a burst is defined as a set of consecutive events such that the IET between each pair of two consecutive events in the burst is less than or equal to $\Delta t$, while the IETs between bursts are greater than $\Delta t$~\cite{Karsai2012Universal}. We call the number of events in each burst as the burst size, and it is denoted by $b$. Any successive bursts separated by some IET, say $\tau$, are supposed to merge when $\Delta t$ increases to become $\tau$. By scanning $\Delta t$ from $\tau_{\rm min}$ to $\tau_{\rm max}$, the burst-tree decomposition method captures such merging process, hence the hierarchical structure of bursts without losing any information in the original event sequence.

\begin{figure*}[!t]
\centering
\includegraphics[width=0.9\textwidth]{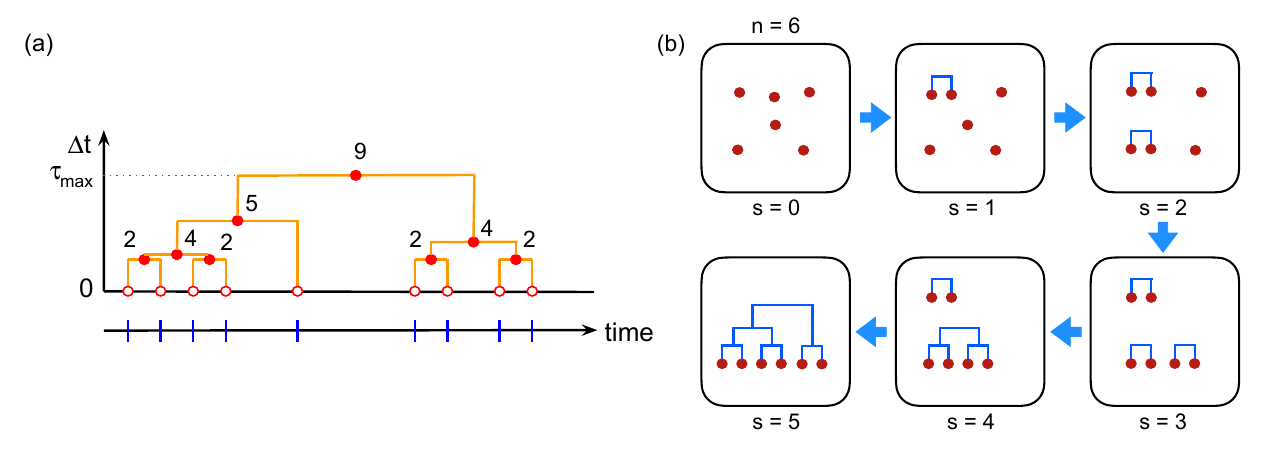}
\caption{(a) Illustration of the burst-tree decomposition method. Vertical blue lines on the time axis in the lower panel represent events. The vertical axis in the upper panel represents a timescale $\Delta t$. Initially, when $\Delta t < \tau_{\rm min}$, each event makes a burst of size one (empty red circles). As $\Delta t$ increases from $\tau_{\rm min}$ to $\tau_{\rm max}$, the bursts are sequentially merged to form bigger bursts (filled red circles). For $\Delta t$ $\ge$ $\tau_{\rm max}$, all events are clustered to form one giant burst. The number next to the merged burst denotes the burst size. (b) Schematic diagram of the stochastic merging process generating an ordinal burst tree with $n=6$. }
\label{fig:burst_tree_fig}
\end{figure*}

We describe how the burst tree $\mathcal{T}$ is constructed by varying the timescale $\Delta t$. Initially, we set $\Delta t$ to be smaller than $\tau_{\rm min}$; each event is considered a burst of size one, depicted as empty red circles in Fig.~\ref{fig:burst_tree_fig}(a). The merging process starts when $\Delta t$ continuously increases from $\tau_{\rm min}$ to $\tau_{\rm max}$. Accordingly, the consecutive bursts are sequentially merged to form bigger bursts, depicted as filled red circles in Fig.~\ref{fig:burst_tree_fig}(a). Finally, the merging process ends when $\Delta t \geq \tau_{\rm max}$, when all events belong to one giant burst of size $n$. Such merging pattern can be visualized by a dendrogram in Fig.~\ref{fig:burst_tree_fig}(a), also called a burst tree; individual events, merged bursts, and the giant burst of size $n$ are represented as leaf nodes, internal nodes, and the root node in the burst tree, respectively. The numbers next to the filled circles in Fig.~\ref{fig:burst_tree_fig}(a) are burst sizes of internal nodes. For example, the burst tree derived from the sample event sequence in Fig.~\ref{fig:template_fig}(a) is shown in Fig.~\ref{fig:template_fig}(b). Then, for any given $\Delta t$, a burst size sequence can be obtained directly from the burst tree, and it is denoted by $\mathcal{B}_{\Delta t}=\{b_1,\ldots, b_m\}$, where $m$ is the number of bursts detected using $\Delta t$. Note that $\sum_{j=1}^m b_j=n$ for any $\Delta t$. Using the burst size sequence $\mathcal{B}_{\Delta t}$, one can characterize the higher-order temporal correlations in terms of the burst size distribution $Q_{\Delta t}(b)$, the burst complexity $C_{\Delta t}$, and the memory coefficient for bursts $M_{\Delta t}$, which will be defined below.

The burst size distribution $Q_{\Delta t}(b)$ obtained from $\mathcal{B}_{\Delta t}$ often shows a heavy tail~\cite{Karsai2012Universal, Karsai2018Bursty, Jo2020Bursttree}, clearly indicating the presence of long-term correlations between a number of consecutive IETs. It is obviously beyond the correlations characterized by $P(\tau)$, $A$, and $M_\tau$. To summarize such higher-order temporal correlations, we propose the burst complexity which quantifies the variation of burst sizes normalized by their mean:
\begin{align}
    C_{\Delta t}\equiv \frac{\sqrt {m + 1} r_b - \sqrt{m - 1}}{(\sqrt{m + 1} - 2) r_b + \sqrt{m - 1}},
    \label{eq:C_deltat}
\end{align}
where $r_b \equiv \sigma_b/\mu_b$ is the coefficient of variation of burst sizes in $\mathcal{B}_{\Delta t}$, with $\sigma_b$ and $\mu_b$ being the standard deviation and the mean of burst sizes, respectively. Similarly to the burstiness measure in Eq.~\eqref{eq:burstiness}, $C_{\Delta t}$ has a value in the range of $[-1,1]$. If all burst sizes are identical, $r_b=0$, hence $C_{\Delta t}=-1$. When all IETs are uncorrelated with each other, the burst size distribution becomes exponential, leading to $C_{\Delta t}\approx 0$. Finally, when IETs are strongly correlated with each other, the complex structure of bursts emerges, as evidenced by the heavy-tailed burst size distributions, hence one may find that $C_{\Delta t}>0$. Note that the number of bursts $m$ decreases with $\Delta t$, and eventually $m$ will decrease to 1; we calculate $C_{\Delta t}$ as long as at least 2 burst sizes are detected. 

Next, we adopt the memory coefficient for bursts $M_{\Delta t}$, quantifying the correlation between two consecutive burst sizes~\cite{Jo2020Bursttree, Birhanu2025Bursttree}:
\begin{align}
    M_{\Delta t} \equiv \frac{1}{m-1}\sum_{j = 1}^{m -1} \frac{(b_j - \mu'_1)(b_{j+1} - \mu'_2)}{\sigma'_1 \sigma'_2},
    \label{eq:M_deltat}
\end{align}
where $\mu'_1$ ($\mu'_2$) and $\sigma'_1$ ($\sigma'_2$) are the average and standard deviation of burst sizes in $\mathcal{B}_{\Delta t}$ excluding $b_m$ ($b_1$), respectively. $M_{\Delta t}$ has a value in the range of $[-1,1]$; the positive $M_{\Delta t}$ implies that bigger (smaller) bursts tend to follow bigger (smaller) bursts, and it is negative for the opposite tendency, while it is around zero for the uncorrelated burst sizes. We calculate $M_{\Delta t}$ as long as at least 3 burst sizes are detected, when both $\sigma'_1$ and $\sigma'_2$ can be calculated. 

We discuss some possible patterns of $C_{\Delta t}$ and $M_{\Delta t}$ regarding burst trees. For very small $\Delta t$, most bursts have the size of one, leading to $C_{\Delta t}\approx -1$ and $M_{\Delta t}\approx 0$. As $\Delta t$ increases, $C_{\Delta t}$ is expected to have non-trivial, positive values, in particular, when the burst size distribution develops a heavy tail. In contrast, one cannot tell about the general tendency of $M_{\Delta t}$ values. Finally, if $\Delta t$ is sufficiently large, there remain only few, big bursts, lowering $C_{\Delta t}$ again. Just because only a few bursts exist, $M_{\Delta t}$ may have extreme values close to $1$ or $-1$.

For the NSR subject 1, we get the empirical results of $Q_{\Delta t}(b)$ for several $\Delta t$s, as well as $C_{\Delta t}$ and $M_{\Delta t}$ for the entire range of $\Delta t$, as shown in Fig.~\ref{fig:template_fig}(e,f). We find that $Q_{\Delta t}(b)$s show heavy tails; using the ordinary least squares linear regression, the power-law exponent $\beta=1.89(6)$ was estimated for $Q_{\Delta t}(b)$ with $\Delta t=700$ ms in the range of $e^2<b<e^6$. In addition, as $\Delta t$ increases, the value of $C_{\Delta t}$ starts from $-1$, then increases to almost $1$, after which it slightly decreases. In the case with $M_{\Delta t}$, it starts from around $0$ before showing a sharp peak at $\Delta t\approx 500$ ms, which is coincident with the timescale when $C_{\Delta t}$ deviates from $-1$. Then $M_{\Delta t}$ tends to increase until $\Delta t\approx 1200$ ms, after which it jumps to $1$ and then to $-1$. Overall $M_{\Delta t}$ has positive values, implying that bigger (smaller) bursts tend to follow bigger (smaller) bursts across multiple timescales~\cite{Jo2017Modeling}. 

It turns out that other subjects also show qualitatively similar patterns of $C_{\Delta t}$ and $M_{\Delta t}$ to those of the NSR subject 1, enabling us to define novel measures characterizing those $C_{\Delta t}$ and $M_{\Delta t}$. First, the increasing behavior of the $C_{\Delta t}$ for small $\Delta t$ can be quantified in terms of a timescale interval, which is defined as 
\begin{align}
    \Delta \equiv \Delta t_2 -\Delta t_1,
    \label{eq:Delta}
\end{align}
where $\Delta t_1$ and $\Delta t_2$ are the timescales satisfying $C_{\Delta t_1}=-0.8$ and $C_{\Delta t_2}=0.5$, respectively. The shorter the timescale interval, the more quickly $C_{\Delta t}$ increases from $-1$ with $\Delta t$. Second, we identify the peak timescale $\Delta t_{\rm peak}$ maximizing $M_{\Delta t}$ within a specific range of $\Delta t$ as
\begin{align}
    \Delta t_{\rm peak} \equiv {\arg\max}_{\tau_{\rm min}\leq \Delta t\leq \Delta t_{\rm upper}} M_{\Delta t},
    \label{eq:Deltat_peak}
\end{align}
where $\Delta t_{\rm upper}=700$~ms is used in our analysis. These numbers $\Delta$ and $\Delta t_{\rm peak}$ will be measured for each subject to be compared across groups of NSR, CHF, and AF.

\subsection{Ordinal burst tree and the burst-merging kernel}\label{subsec:kernel}

Following the original burst-tree decomposition method, we assume that the burst tree is binary, implying that each merged burst consists of only two bursts. In reality, the merged bursts can be made of more than two bursts, which however is not very common~\cite{Jo2020Bursttree}. Let us consider two successive bursts, indexed by $v$ (earlier burst) and $w$ (later burst), that are separated by an IET $\tau$. At the moment of $\Delta t=\tau$, bursts $v$ and $w$ merge together to become another burst indexed by $u$. Then bursts $u$, $v$, and $w$ are called a parent, a left child, and a right child, and their sizes have the relation as $b_u=b_v+b_w$, where $b_\alpha$ is the burst size of $\alpha$ for $\alpha\in\{u,v,w\}$. Denoting the IET separating $v$ and $w$ by $\hat\tau_u$, we sort those IETs in a descending way. Then the index $u$ is determined as the rank of $\hat\tau_u$ in the ordered IETs; e.g., since the root node is associated with $\tau_{\rm max}$ and $\hat\tau_1=\tau_{\rm max}$, the index of the root node is $1$. Finally, one can represent each merged burst by a tuple $(u,v,w,\hat\tau_u)$. Therefore, the burst tree $\mathcal{T}$ is fully described by the set of these tuples $\{(u,v,w,\hat\tau_u)\}$. $\mathcal{T}$ can be decomposed into the ordinal burst tree $\mathcal{G}\equiv \{(u,v,w)\}$ and the set of IETs $\{\hat\tau_u\}$, which is equivalent to $P(\tau)$ in Eq.~\eqref{eq:Ptau_define}. It is remarkable that the ordinal burst tree is separable from the IET distribution, hence it can be associated with an arbitrary IET distribution.

Although the ordinal burst tree $\mathcal{G}$ carries the detailed information on the given event sequence, one often needs to summarize it for the simpler representation. For this, from the observed $\mathcal{G}$ we estimate the burst-merging kernel $K(b,b')$ with $b$ and $b'$ denoting burst sizes of the left and right children in the burst tree, respectively. Kernels are versatile tools used in various fields. In network science kernels guide for shaping the structure of growing networks over time~\cite{Barabasi2016Network, Pham2015PAFit, Barabasi1999Emergence, Krapivsky2000Connectivity, Newman2001Clustering, Jeong2003Measuring, Sheridan2012Measuring, Kunegis2013Preferential}, while in the coagulation process kernels determine how likely two existing clusters are chosen for coalescence~\cite{Stockmayer1943Theory, Lushnikov1973Evolution, White1982Form, Hendriks1983Coagulation, Aldous1999Deterministic, Lee2001Survey, Leyvraz2003Scaling, Leyvraz2005Rigorous, Wattis2006Introduction, Birhanu2025Bursttree}. In our work, we consider the observed $\mathcal{G}$ as an instance of the underlying stochastic merging process driven by the burst-merging kernel~\cite{Jo2020Bursttree, Birhanu2025Bursttree, Birhanu2025Maximum}. To describe the stochastic merging process, let us denote an auxiliary time step by $s$, which counts the number of merges and has nothing to do with the real time. If the burst-merging kernel is given, the ordinal burst tree can be generated using the kernel, as depicted in Fig.~\ref{fig:burst_tree_fig}(b) for the case with $n=6$. Initially at $s=0$, one has $n$ individual events. At each time step $s$, there are exactly $n-s$ bursts. Among them two bursts of sizes $b$ and $b'$ are randomly chosen with a probability proportional to $K(b,b')$, and they merge together to become another burst of size $b+b'$. The burst of size $b$ ($b'$) can be randomly assigned either the left (right) child or the right (left) child of the parent node. This process is repeated until $s=n-1$, when all events belong to one giant burst. 

For the estimation of the burst-merging kernel from the data, we calculate the burst size distribution after $s$ merges, which is denoted by $Q_s(b)$. We also count the number of merges of burst sizes $b$ and $b'$ at each time step $s$, and denote it by $m_{sbb'}$; by definition of $s$, $m_{sbb'}=0,1$ for each $s$. Using these two quantities $Q_s(b)$ and $m_{sbb'}$ for $s=0,\ldots, n-1$ and $b,b'=1,\ldots,n$, we estimate a matrix of $n\times n$ elements for the burst-merging kernel, denoted by \textbf{K}, by means of the maximum likelihood method~\cite{Birhanu2025Maximum}. We initialize the kernel as $K_0(b,b')=1$ for $1\leq b,b'\leq n$, and then iterate the following equation using $Q_s(b)$ and $m_{sbb'}$:
\begin{align}
    K_i(b,b')=\dfrac{\sum_{s=1}^{n-1}m_{sbb'}}
    {\sum_{s=1}^{n-1} \dfrac{Q_{s-1}(b)Q_{s-1}(b')}{\sum_{k,k'=1}^n Q_{s-1}(k)Q_{s-1}(k')K_{i-1}(k,k')}}
    \label{eq:K_bb'_iter}
\end{align}
for $i=1,2,\ldots$. The iteration is repeated until the following condition is satisfied:
\begin{align}
    \frac{|l(\textbf{K}_i)-l(\textbf{K}_{i-1})|}{|l(\textbf{K}_{i-1})|+1}\leq \epsilon,
    \label{eq:convergence}
\end{align}
where $l(\textbf{K})$ is the log-likelihood of the kernel defined as follows:
\begin{align}
    l(\textbf{K}) &\equiv \sum_{s=1}^{n-1}\sum_{b,b'=1}^n m_{sbb'}\ln K(b,b')\notag\\
    &- \sum_{s=1}^{n-1}\ln \left[ \sum_{k,k'=1}^n Q_{s-1}(k)Q_{s-1}(k')K(k,k')\right].
    \label{eq:l_K}
\end{align}
$\epsilon$ is a convergence parameter that takes a sufficiently small positive value; we use $\epsilon = 10^{-4}$ throughout the paper. Finally, we normalize the estimated kernel to compare them across subjects:
\begin{align}
    \tilde K(b,b') \equiv  \dfrac{K(b,b')}{\sum_{b,b' = 1}^{50}K(b,b')}.
    \label{eq:normalized_kernel}
\end{align}
We will denote the normalized kernel by $K(b,b')$ without the tilde for clearer presentation of our results hereafter. The estimated kernel of the NSR subject 1 is shown in Fig.~\ref{fig:template_fig}(c), where we observe $K(b,b')$ is an overall increasing function of $b$ and $b'$. It implies that bigger bursts are more likely to be chosen for the merge than smaller bursts. This is why the burst size distribution develops a heavy tail and the consecutive burst sizes are positively correlated with each other. In this sense, the burst-merging kernel is of utmost importance to understand the higher-order temporal correlations. 

Yet, the burst-merging kernel $K(b,b')$ is hard to compare across individuals and groups, leading us to introduce two kinds of cross sections of $K(b,b')$, namely, the principal and the secondary diagonals, as depicted by dashed lines in Fig.~\ref{fig:template_fig}(c). These diagonals are respectively defined as
\begin{align}
    K_{1}(b) \equiv K(b,b),\ 
    K_{2}(b) \equiv K\left( b,\frac{1500}{b} \right).
    \label{eq:diagonals}
\end{align}
Note that the constant of $1500$ for $K_2(b)$ has been chosen to maximize the range of $b$ for a given $K(b,b')$. Figure~\ref{fig:template_fig}(g) shows these diagonals $K_1(b)$ and $K_2(b)$. We find that $K_1(b)$ overall increases with $b$, while $K_2(b)$ shows an overall symmetric behavior with respect to $b=\sqrt{1500}\approx 39$, implying the time-reversal symmetry in terms of the burst sizes.

In conclusion, we propose the analysis framework for a given event sequence by characterizing the entire orders of temporal correlations in the event sequence. We begin by measuring the IET distribution $P(\tau)$, the burstiness measure $A$, and the memory coefficient $M_\tau$, then after mapping the event sequence to the burst tree we measure the burst size distribution $Q_{\Delta t}(b)$, the burst complexity $C_{\Delta t}$, and the memory coefficient for bursts $M_{\Delta t}$ for the entire range of $\Delta t$. Finally, we estimate the burst-merging kernel $K(b,b')$ summarizing the ordinal burst tree and take the principal and secondary cross sections of $K(b,b')$ as $K_1(b)$ and $K_2(b)$, making it easier to compare these quantities across subjects and groups.

In addition, to compare the distributions of quantities, such as $A$ and $M_\tau$, across groups of NSR, CHF, and AF, we conduct the two-sample Kolmogorov-Smirnov test comparing those distributions. By using the \texttt{ks\_2samp} function from the \texttt{scipy.stats} library in Python, we calculate the p-values and report them as appropriate.

\subsection{Classification of individuals using SVM}\label{subsec:class}

Finally, we adopt the support vector machine (SVM)~\cite{Cortes1995Supportvector} to test if one can classify individual subjects into three groups of NSR, CHF, and AF. Since the SVM has originally been for binary classification, we employ the one-vs-rest method for multi-class classification using the \texttt{sklearn} library in Python. As for input features, we use the burstiness measure $A$ in Eq.~\eqref{eq:burstiness}, memory coefficient $M_{\tau}$ in Eq.~\eqref{eq:memory}, timescale interval $\Delta$ in Eq.~\eqref{eq:Delta}, and peak timescale $\Delta t_{\rm peak}$ in Eq.~\eqref{eq:Deltat_peak}. Since values of these features are on different scales, we normalize them in terms of Z-score. We employ the radial basis function kernel and perform grid search using $5$-fold cross-validation to determine hyperparameters, i.e., the regularization parameter that controls the trade-off between low training and testing errors, and the kernel parameter that defines the influence of a single training example, with higher values leading to more complex decision boundaries. We perform the classification $100$ times or runs; in each run, we randomly choose $80\%$ of samples for the train set, and the rest $20\%$ as the test set.

As a result of the classification using the SVM, one obtains the confusion matrix of $3\times 3$ elements; each element $n_{ij}$ for $i,j\in\{0,1,2\}$ denotes the number of subjects who belong to the class $i$ but are predicted to belong to the class $j$. Here the classes indexed by $0,1,2$ correspond to the groups of NSR, CHF, and AF, respectively. For each class $i$ we consider $n_{ii}$ as true positive (TP), $n_{ij}$ with $j\neq i$ as false negative (FN), $n_{ji}$ with $j\neq i$ as false positive (FP), and $n_{jk}$ with $j,k\neq i$ as true negative (TN). Using these numbers, we calculate the performance metrics, namely, sensitivity and specificity for each class $i\in\{0,1,2\}$ as well as the overall accuracy and balanced accuracy as follows~\cite{Tharwat2021Classification, Grandini2020Metrics}:
\begin{align}
    &\text{Sensitivity}_i \equiv \frac{n_{ii}}{\sum_{j}n_{ij}}\label{eq:sens_af},\\
    &\text{Specificity}_i \equiv \frac{\sum_{j\neq i}\sum_{k\neq i} n_{jk}}{\sum_{j\neq i}\sum_{k} n_{jk}}\label{eq:spec_af},\\
    &\text{Overall Accuracy}\equiv \frac{\sum_{i}n_{ii}}{\sum_{i}\sum_j n_{ij}}\label{eq:oa},\\
    &\text{Balanced Accuracy}\equiv \frac{1}{3}\sum_{i}\frac{n_{ii}}{\sum_{j} n_{ij}}\label{eq:ba}.
\end{align}
The sensitivity measures the ability of the classifier, which is the multi-class SVM in our case, to correctly identify individual subjects with a positive case, while the specificity indicates the ability of the classifier to correctly identify those with a negative case. Note that the balanced accuracy is equivalent to the average of the sensitivities of all classes. The implementation codes for our analysis framework are available in the GitHub repository~\cite{Birhanu2025Codesa}. 

\begin{figure*}[!t]
    \centering
    \includegraphics[width=0.9\textwidth]{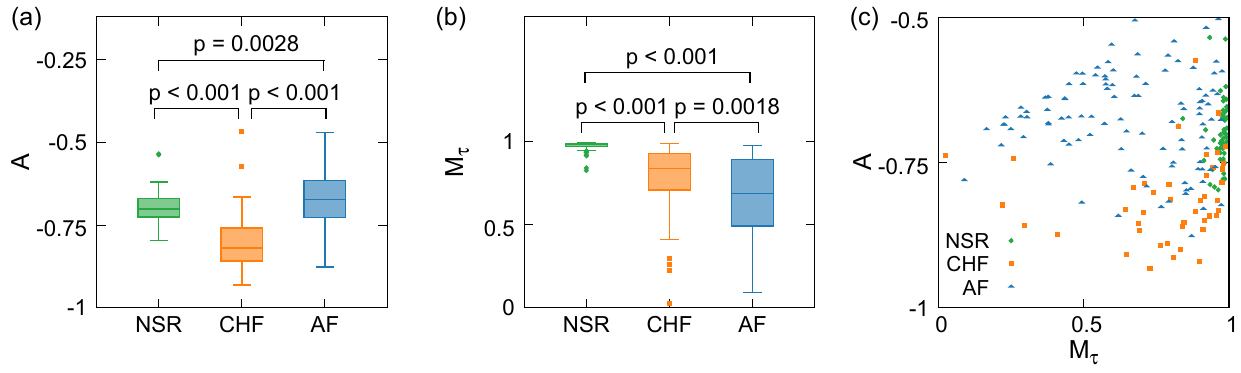}
    \caption{Empirical results of the burstiness measure $A$ in Eq.~\eqref{eq:burstiness} and the memory coefficient $M_\tau$ in Eq.~\eqref{eq:memory}. (a) Box plots of $A$ values for NSR, CHF, and AF groups. (b) Box plots of $M_{\tau}$ values for the same three groups. In panels (a) and (b), we include the p-values for the two-sample Kolmogorov-Smirnov test comparing distributions of $A$ and $M_\tau$. (c) Scatter plot of individual results in the $(M_\tau,A)$ space, highlighting different regions occupied by different groups.}
    \label{fig:burstiness_memory_of_iet}
\end{figure*}

\begin{figure*}[!t]
    \centering
    \includegraphics[width=\textwidth]{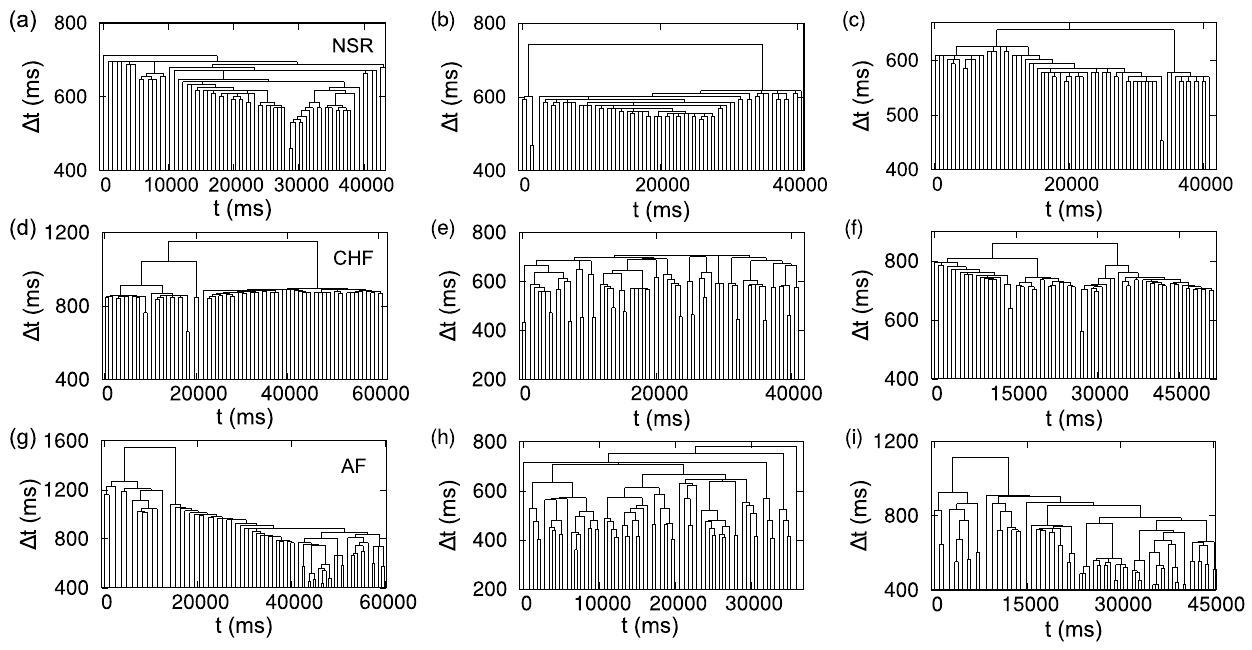}
    \caption{Visualization of burst trees derived from event sequences for three selected subjects in each of NSR, CHF, and AF groups (top to bottom). In all cases, $70$ consecutive events are sampled for visualization. Note that the range of $\Delta t$ in the vertical axis is different across panels.}
    \label{fig:multiple_burst_tree}
\end{figure*}

\begin{figure*}[!t]
    \centering
    \includegraphics[width=\textwidth]{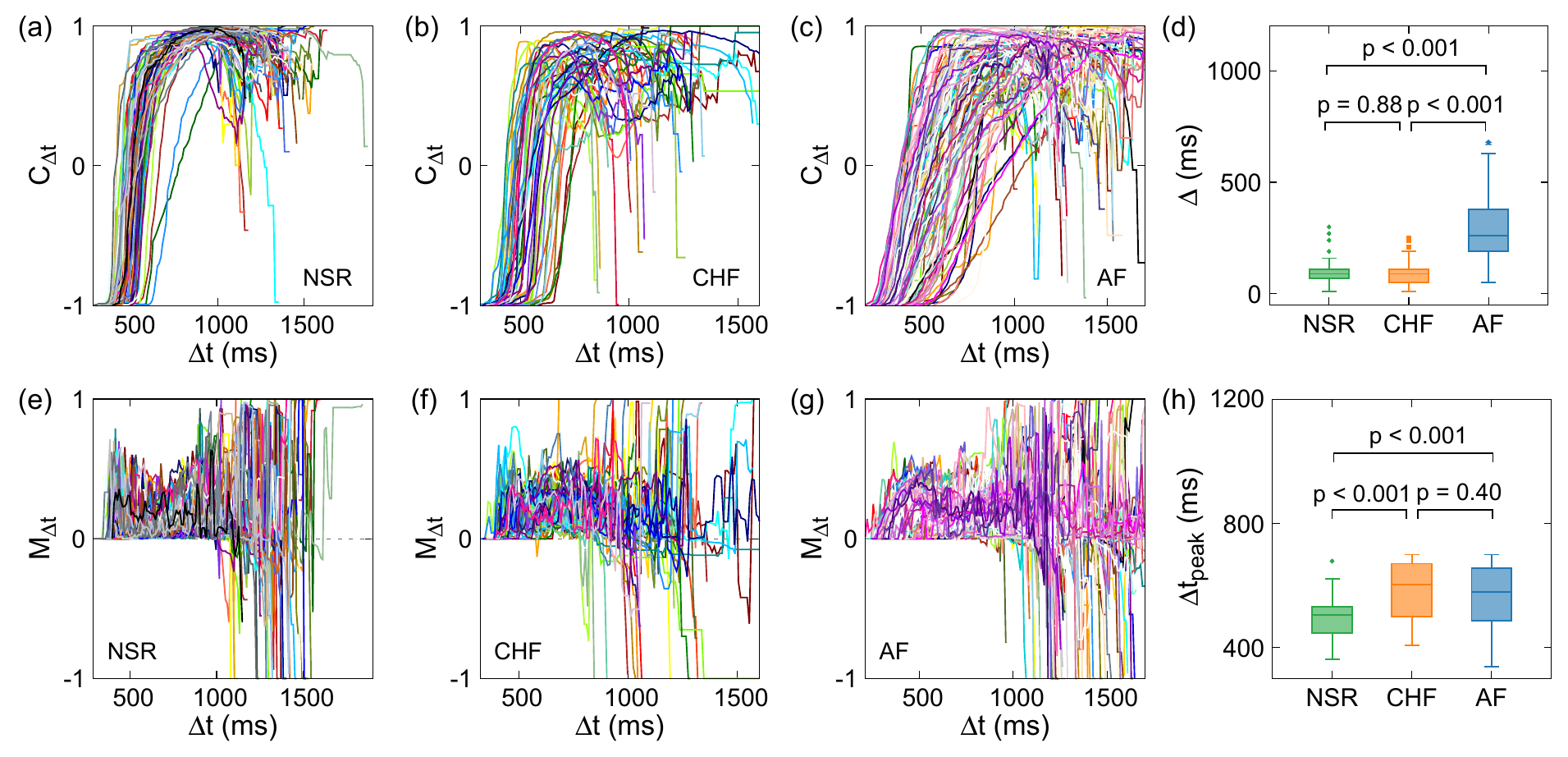}
    \caption{Empirical results of the burst complexity $C_{\Delta t}$ in Eq.~\eqref{eq:C_deltat} and the memory coefficient for bursts $M_{\Delta t}$ in Eq.~\eqref{eq:M_deltat}. (a--c) The curves of $C_{\Delta t}$ for the entire range of $\Delta t$ in each of NSR, CHF, and AF groups. (d) Box plots of the timescale interval, defined as $\Delta\equiv \Delta t_2 - \Delta t_1$ in Eq.~\eqref{eq:Delta}, where $\Delta t_1$ and $\Delta t_2$ are the timescales satisfying $C_{\Delta t_1}=-0.8$ and $C_{\Delta t_2}=0.5$. We include the p-values for the two-sample Kolmogorov-Smirnov test comparing distributions of $\Delta$. (e--g) The curves of $M_{\Delta t}$ for the entire range of $\Delta t$ in each of NSR, CHF, and AF groups. The horizontal dotted lines at $M_{\Delta t}=0$ are to guide the eyes. (h) Box plots of the peak timescale, defined as $\Delta t_{\rm peak} \equiv {\arg\max}_{\tau_{\rm min}\leq \Delta t\leq \Delta t_{\rm upper}} M_{\Delta t}$ in Eq.~\eqref{eq:Deltat_peak} with $\Delta t_{\rm upper}=700$~ms. We include the p-values for the two-sample Kolmogorov-Smirnov test comparing distributions of $\Delta t_{\rm peak}$.
    }   
    \label{fig:cv_memory_of_burst_size}
\end{figure*}

\begin{figure*}[!t]
    \centering
    \includegraphics[width=0.9\textwidth]{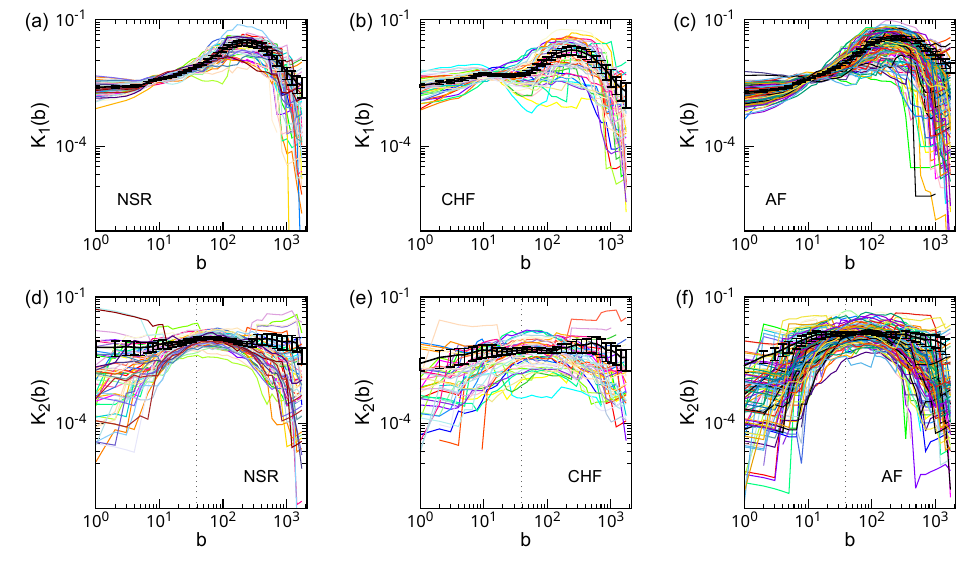}
    \caption{Empirical results of the principal and secondary cross sections $K_1(b)$ (a--c) and $K_2(b)$ (d--f) in Eq.~\eqref{eq:diagonals} derived from the estimated burst-merging kernel $K(b,b')$. The black curve in each panel shows the average of the curves with the standard error. The vertical dotted lines at $b\approx 39$ in panels (d--f) are to guide the eyes.}
    \label{fig:diagonal_kernel}
\end{figure*}

\begin{figure}[!t]
    \centering
    \includegraphics[width=0.7\columnwidth]{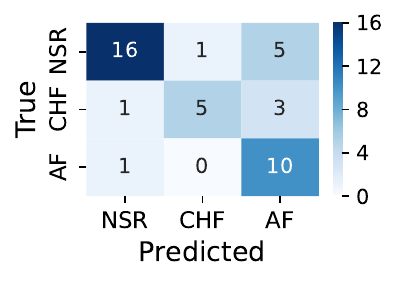}
    \caption{Confusion matrix for groups of NSR, CHF, and AF from the single run of the SVM method.}
    \label{fig:confusion_matrix}
\end{figure}

\section{Results}\label{sec:result}

We apply our analysis framework proposed in the previous section to the heartbeat time series of 207 subjects in three groups of the normal sinus rhythm (NSR), the congestive heart failure (CHF), and the atrial fibrillation (AF) to see whether there are significant differences between these groups in terms of temporal correlations of all relevant orders.

For each subject, we first estimate the values of the burstiness measure $A$ in Eq.~\eqref{eq:burstiness} and the memory coefficient $M_\tau$ in Eq.~\eqref{eq:memory}. Figure~\ref{fig:burstiness_memory_of_iet}(a,b) show the distributions of those results in the form of box plots. We observe that $A<-0.5$ for all subjects as the heartbeats are mostly regular rather than bursty; the range of $A$ values for the NSR group is higher than that for the CHF group, implying that the subjects with CHF show more regular heartbeats than the healthy subjects. The values of $A$ for the AF group are most dispersed, though their median is closer to that for the NSR group than for the CHF group. The differences between groups get more pronounced in the distribution of $M_\tau$; the values of $M_\tau$ are all very close to $1$ for the NSR group, while those for the CHF group are obviously smaller than those of the NSR group. The values of $M_\tau$ for the AF group are overall smallest and again most dispersed, while overlapping with those of the CHF group. Our observations are supported by the p-values estimated between groups using the KS test. Finally, we generate a scatter plot of the results in the $(M_\tau,A)$ space in Fig.~\ref{fig:burstiness_memory_of_iet}(c) to see if the groups can be distinguished from each other. Three groups occupy overall different regions but still with some overlaps.

Next, we map the event sequence of each subject onto the burst tree. For the qualitative comparison, we visualize burst trees derived from three selected subjects for each group, each of which contains exactly 70 consecutive events, as shown in Fig.~\ref{fig:multiple_burst_tree}. Burst trees for the NSR and CHF groups appear to have more regular IETs in the shorter range of IETs, hence more regular structure, than those for the AF group. 

To quantify such (ir)regularities in burst trees, we measure for each subject the burst complexity $C_{\Delta t}$ in Eq.~\eqref{eq:C_deltat} and the memory coefficient for bursts $M_{\Delta t}$ in Eq.~\eqref{eq:M_deltat} for the entire range of timescales $\tau_{\rm min}\leq \Delta t\leq \tau_{\rm max}$. We plot the curves of $C_{\Delta t}$ ($M_{\Delta t}$) for all subjects in each of NSR, CHF, and AF groups in Fig.~\ref{fig:cv_memory_of_burst_size}(a--c) [Fig.~\ref{fig:cv_memory_of_burst_size}(e--g)]. The $C_{\Delta t}$ curves for the NSR group show more coherent behaviors than those for the CHF group, whereas those for the AF group show most diverse behaviors. Most $C_{\Delta t}$ curves for the NSR group get very close to $1$. On the other hand, for the CHF and AF groups only a few subjects show $C_{\Delta t}$ approaching $1$, and $C_{\Delta t}$ curves are much more fluctuating than those for the NSR group. This observation may indicate that the NSR time series have more complex structure of bursts than other groups. For the quantitative comparison of $C_{\Delta t}$ curves between groups, we focus on the increasing behavior of $C_{\Delta t}$ from $-1$ to a large positive value, e.g., $0.5$, in terms of the timescale interval $\Delta$ in Eq.~\eqref{eq:Delta}. As shown in Fig.~\ref{fig:cv_memory_of_burst_size}(d), the timescale intervals for the NSR and CHF groups are much smaller than those for the AF group, which is also consistent with the observations for burst trees in Fig.~\ref{fig:multiple_burst_tree}. We perform the sensitivity analysis using different values of $C_{\Delta t_1}$ and $C_{\Delta t_2}$ to find that our conclusions are robust with respect to such values as depicted in Table~\ref{tab:C_p-value}.

\begin{table}[!t]
   \centering
      \begin{tabular}{llccc} \hline \hline
       $C_{\Delta t_1}$  & $C_{\Delta t_2}$ & NSR vs. CHF & NSR vs. AF & CHF vs. AF\\ \hline
       $ -0.9$ & $0.4$ & $0.920$ & $4.92 \times 10^{-24}$ & $ 2.30 \times 10^{-17}$\\
       $ -0.9$  & $0.45$ & $0.913$ & $4.92 \times 10^{-24}$  & $4.43  \times 10^{-18}$\\
       $ -0.9$ & $0.5$ & $0.734$  & $4.92 \times 10^{-24}$ &$ 1.44 \times 10^{-16}$ \\
       $-0.8 $  & $0.4$ & $0.903$ & $7.90 \times 10^{-23}$ & $ 2.55 \times 10^{-16}$\\
       $-0.8$   & $0.45$ & $0.977$ & $8.85 \times 10^{-23}$ & $ 2.00 \times 10^{-16}$\\
       $-0.8 $ & $0.5 $ & $0.878$ & $3.77 \times 10^{-22}$ &$ 1.18 \times 10^{-15}$ \\
       $-0.7 $  & $0.4$ & $0.548$ & $3.40 \times 10^{-22}$ & $3.38  \times 10^{-17}$\\
       $-0.7 $  & $0.45$ & $0.940$ & $9.35 \times 10^{-22}$ & $6.93  \times 10^{-18}$\\
       $-0.7$ & $0.5$ & $0.748$ & $1.18 \times 10^{-21}$ & $1.01  \times 10^{-16}$\\ \hline \hline
    \end{tabular}
    \caption{Sensitivity analysis with respect to the choice of $C_{\Delta t_1}$ and $C_{\Delta t_2}$ in terms of p-values estimated using each pair of groups of NSR, CHF, and AF.}
    \label{tab:C_p-value}
\end{table}

\begin{table}[!t]
    \centering
    \begin{tabular}{cccc}\hline \hline 
    $\Delta t_{\rm upper}$ (ms) &  NSR vs. CHF & NSR vs. AF & CHF vs. AF\\ \hline 
    600 & $1.17 \times 10^{-4}$  & $1.32 \times 10^{-4}$ & $0.560$\\
    650 & $6.89 \times 10^{-6}$  & $1.17 \times 10^{-4}$ & $0.267$\\
    700 & $1.60 \times 10^{-6}$ & $1.48 \times 10^{-5}$ & $0.395$\\
    750 & $8.49 \times 10^{-8}$ & $1.17 \times 10^{-6}$ & $0.246$\\
    800 & $3.66 \times 10^{-5}$ & $4.94 \times 10^{-4}$ & $0.302$\\
    \hline \hline 
    \end{tabular}
    \caption{Sensitivity analysis with respect to the choice of the upper bound of the range of timescales, i.e., $\Delta t_{\rm upper}$, in terms of p-values estimated using each pair of groups of NSR, CHF, and AF.
    }    
    \label{tab:M_p-value}
\end{table}

We also find in Fig.~\ref{fig:cv_memory_of_burst_size}(e--g) that in most cases $M_{\Delta t}$ initially increases from $\approx 0$ at $\Delta t=\tau_{\rm min}$ and then fluctuates as the timescale $\Delta t$ increases before ending up with either $1$ or $-1$, when there are only a few bursts left. Based on the observation that the peaks of $M_{\Delta t}$ curves are located at different timescales, we focus on the peak timescale $\Delta t_{\rm peak}$ that maximizes the value of $M_{\Delta t}$ for small $\Delta t$ as defined in Eq.~\eqref{eq:Deltat_peak} for each subject; we compare the distributions of $\Delta t_{\rm peak}$ values across groups to find that $\Delta t_{\rm peak}$ values for the NSR group are overall smaller than those for the CHF and AF groups as shown in Fig.~\ref{fig:cv_memory_of_burst_size}(h). We perform the sensitivity analysis using different values of $\Delta t_{\rm upper}$ to find that our conclusions are robust with respect to such values as depicted in Table~\ref{tab:M_p-value}.

Next we measure the principal and secondary diagonal cross sections $K_1(b)$ and $K_2(b)$ in Eq.~\eqref{eq:diagonals} of the estimated burst-merging kernel for each subject. The results are depicted in Fig.~\ref{fig:diagonal_kernel}. To reduce noise, we have smoothened the curves by applying a moving average with window size of $9$. We observe that $K_1(b)$ is an overall increasing function of $b$ for the range of $1\leq b\leq 300$ in all groups, while the increasing patterns are slightly different across groups. Black solid curves denote the average of curves in each panel. In all cases, $K_2(b)$ appears to be overall symmetric, while they are also different across groups. Studying such differences may deepen the understanding of heartbeat time series of people in various heart conditions, which we leave for future work.

Finally, we observe that each of four measures, namely, burstiness measure $A$ in Eq.~\eqref{eq:burstiness}, memory coefficient $M_{\tau}$ in Eq.~\eqref{eq:memory}, timescale interval $\Delta$ in Eq.~\eqref{eq:Delta}, and peak timescale $\Delta t_{\rm peak}$ in Eq.~\eqref{eq:Deltat_peak} has been able to distinguish one group from the others among three groups of NSR, CHF, and AF. Thus, we expect to distinguish those groups better when using all of the mentioned four measures simultaneously than when using one of them only. For this, we adopt the support vector machine (SVM) method as described in the Materials and Methods section. The confusion matrix resulting from the SVM method for the single run is shown in Fig.~\ref{fig:confusion_matrix}. The averages and standard deviations of the performance metrics in Eqs.~\eqref{eq:sens_af}--\eqref{eq:ba} over $100$ different runs of the SVM method are presented in Table~\ref{tab:sens_spec}. We achieve the overall accuracy of $80\%$ and the balanced accuracy of $78\%$ on average. The balanced accuracy of $78\%$ turns out to be higher than the reported value of $66\%$ when using the DFA exponent as an input feature, while it is lower than $82\%$ when using the DDFA exponent~\cite{Pukkila2025Detection}. Our classification results show the possibility that our analysis framework for the higher-order temporal correlations might be useful for the classification and detection of subjects with heart diseases.

\begin{table}[!t]
    \centering
    \begin{tabular}{ccc}
    \hline \hline
    Group & Sensitivity & Specificity\\
    \hline
    NSR  & $0.83 \pm 0.09$ & $0.87 \pm 0.09$ \\
    CHF & $0.69 \pm 0.15$  & $0.93 \pm 0.04$ \\
    AF & $0.83 \pm 0.12$ & $0.88 \pm 0.06$ \\ \hline
    \multicolumn{2}{c}{Overall Accuracy} & $0.80 \pm 0.06$\\
    \multicolumn{2}{c}{Balanced Accuracy} & $0.78 \pm 0.07$ \\
    \hline \hline
    \end{tabular}
    \caption{Estimated values of the sensitivity in Eq.~\eqref{eq:sens_af} and the specificity in Eq.~\eqref{eq:spec_af} for each group of NSR, CHF, and AF; averages and standard deviations are obtained from $100$ different runs of the SVM method. We also present the estimated values of overall and balanced accuracies in Eqs.~\eqref{eq:oa} and~\eqref{eq:ba} from the same $100$ runs.}
    \label{tab:sens_spec}
\end{table}

\section{Conclusion}\label{sec:conclusion}

We have proposed a novel analysis framework for high-order temporal correlations in event sequences mainly based on the burst-tree decomposition method~\cite{Jo2020Bursttree}; we have successfully applied our framework to human heartbeat time series of 207 people from different groups of the normal sinus rhythm (NSR), the congestive heart failure (CHF), and the atrial fibrillation (AF)~\cite{PhysioBank}. By measuring various existing and novel measures from each subject's time series, we could compare the statistical properties of such time series across groups. In particular, we explore the possibility that higher-order quantities derived from the burst-tree structure, such as the burst complexity and memory coefficient for bursts, can be used to distinguish different groups of people. We adopt one of commonly used machine learning techniques, i.e., the support vector machine, to achieve the reasonable accuracy for the classification task. In sum, our work highlights the potential of the burst-tree decomposition method and our analysis framework based on the method to improve the understanding of complex physiological time series data, paving the way for further research in the fields of complex systems and nonlinear time series analysis.

\begin{acknowledgments}
T.B. and H.-H.J. acknowledge financial support by the National Research Foundation of Korea (NRF) grant funded by the Korea government (MSIT) (No. 2022R1A2C1007358).
\end{acknowledgments}

%apsrev4-2.bst 2019-01-14 (MD) hand-edited version of apsrev4-1.bst
%Control: key (0)
%Control: author (8) initials jnrlst
%Control: editor formatted (1) identically to author
%Control: production of article title (0) allowed
%Control: page (0) single
%Control: year (1) truncated
%Control: production of eprint (0) enabled
%

%\bibliography{h2jo-papers}% Produces the bibliography via BibTeX.

\begin{thebibliography}{62}%
\makeatletter
\providecommand \@ifxundefined [1]{%
 \@ifx{#1\undefined}
}%
\providecommand \@ifnum [1]{%
 \ifnum #1\expandafter \@firstoftwo
 \else \expandafter \@secondoftwo
 \fi
}%
\providecommand \@ifx [1]{%
 \ifx #1\expandafter \@firstoftwo
 \else \expandafter \@secondoftwo
 \fi
}%
\providecommand \natexlab [1]{#1}%
\providecommand \enquote  [1]{``#1''}%
\providecommand \bibnamefont  [1]{#1}%
\providecommand \bibfnamefont [1]{#1}%
\providecommand \citenamefont [1]{#1}%
\providecommand \href@noop [0]{\@secondoftwo}%
\providecommand \href [0]{\begingroup \@sanitize@url \@href}%
\providecommand \@href[1]{\@@startlink{#1}\@@href}%
\providecommand \@@href[1]{\endgroup#1\@@endlink}%
\providecommand \@sanitize@url [0]{\catcode `\\12\catcode `\$12\catcode
  `\&12\catcode `\#12\catcode `\^12\catcode `\_12\catcode `\%12\relax}%
\providecommand \@@startlink[1]{}%
\providecommand \@@endlink[0]{}%
\providecommand \url  [0]{\begingroup\@sanitize@url \@url }%
\providecommand \@url [1]{\endgroup\@href {#1}{\urlprefix }}%
\providecommand \urlprefix  [0]{URL }%
\providecommand \Eprint [0]{\href }%
\providecommand \doibase [0]{https://doi.org/}%
\providecommand \selectlanguage [0]{\@gobble}%
\providecommand \bibinfo  [0]{\@secondoftwo}%
\providecommand \bibfield  [0]{\@secondoftwo}%
\providecommand \translation [1]{[#1]}%
\providecommand \BibitemOpen [0]{}%
\providecommand \bibitemStop [0]{}%
\providecommand \bibitemNoStop [0]{.\EOS\space}%
\providecommand \EOS [0]{\spacefactor3000\relax}%
\providecommand \BibitemShut  [1]{\csname bibitem#1\endcsname}%
\let\auto@bib@innerbib\@empty
%</preamble>
\bibitem [{\citenamefont {Sayama}(2015)}]{Sayama2015Introduction}%
  \BibitemOpen
  \bibfield  {author} {\bibinfo {author} {\bibfnamefont {H.}~\bibnamefont
  {Sayama}},\ }\href@noop {} {\emph {\bibinfo {title} {Introduction to the
  Modeling and Analysis of Complex Systems}}},\ \bibinfo {edition} {deluxe
  color ed}\ ed.\ (\bibinfo  {publisher} {Open SUNY Textbooks, Milne Library},\
  \bibinfo {address} {Geneseo, NY},\ \bibinfo {year} {2015})\BibitemShut
  {NoStop}%
\bibitem [{\citenamefont {Barab{\'a}si}(2005)}]{Barabasi2005Origin}%
  \BibitemOpen
  \bibfield  {author} {\bibinfo {author} {\bibfnamefont {A.-L.}\ \bibnamefont
  {Barab{\'a}si}},\ }\bibfield  {title} {\bibinfo {title} {The origin of bursts
  and heavy tails in human dynamics},\ }\href
  {https://doi.org/10.1038/nature03459} {\bibfield  {journal} {\bibinfo
  {journal} {Nature}\ }\textbf {\bibinfo {volume} {435}},\ \bibinfo {pages}
  {207} (\bibinfo {year} {2005})}\BibitemShut {NoStop}%
\bibitem [{\citenamefont {Karsai}\ \emph {et~al.}(2018)\citenamefont {Karsai},
  \citenamefont {Jo},\ and\ \citenamefont {Kaski}}]{Karsai2018Bursty}%
  \BibitemOpen
  \bibfield  {author} {\bibinfo {author} {\bibfnamefont {M.}~\bibnamefont
  {Karsai}}, \bibinfo {author} {\bibfnamefont {H.-H.}\ \bibnamefont {Jo}},\
  and\ \bibinfo {author} {\bibfnamefont {K.}~\bibnamefont {Kaski}},\
  }\href@noop {} {\emph {\bibinfo {title} {Bursty {{Human Dynamics}}}}}\
  (\bibinfo  {publisher} {Springer International Publishing},\ \bibinfo
  {address} {Cham},\ \bibinfo {year} {2018})\BibitemShut {NoStop}%
\bibitem [{\citenamefont {Jo}\ and\ \citenamefont
  {Hiraoka}(2023)}]{Jo2023Bursty}%
  \BibitemOpen
  \bibfield  {author} {\bibinfo {author} {\bibfnamefont {H.-H.}\ \bibnamefont
  {Jo}}\ and\ \bibinfo {author} {\bibfnamefont {T.}~\bibnamefont {Hiraoka}},\
  }\bibfield  {title} {\bibinfo {title} {Bursty {{Time Series Analysis}} for
  {{Temporal Networks}}},\ }in\ \href
  {https://doi.org/10.1007/978-3-031-30399-9_9} {\emph {\bibinfo {booktitle}
  {Temporal {{Network Theory}}}}},\ \bibinfo {editor} {edited by\ \bibinfo
  {editor} {\bibfnamefont {P.}~\bibnamefont {Holme}}\ and\ \bibinfo {editor}
  {\bibfnamefont {J.}~\bibnamefont {Saram{\"a}ki}}}\ (\bibinfo  {publisher}
  {Springer International Publishing},\ \bibinfo {address} {Cham},\ \bibinfo
  {year} {2023})\ \bibinfo {edition} {2nd}\ ed.,\ pp.\ \bibinfo {pages}
  {165--183}\BibitemShut {NoStop}%
\bibitem [{\citenamefont {Wheatland}\ \emph {et~al.}(1998)\citenamefont
  {Wheatland}, \citenamefont {Sturrock},\ and\ \citenamefont
  {McTiernan}}]{Wheatland1998WaitingTime}%
  \BibitemOpen
  \bibfield  {author} {\bibinfo {author} {\bibfnamefont {M.~S.}\ \bibnamefont
  {Wheatland}}, \bibinfo {author} {\bibfnamefont {P.~A.}\ \bibnamefont
  {Sturrock}},\ and\ \bibinfo {author} {\bibfnamefont {J.~M.}\ \bibnamefont
  {McTiernan}},\ }\bibfield  {title} {\bibinfo {title} {The {{Waiting-Time
  Distribution}} of {{Solar Flare Hard X-Ray Bursts}}},\ }\href
  {https://doi.org/10.1086/306492} {\bibfield  {journal} {\bibinfo  {journal}
  {The Astrophysical Journal}\ }\textbf {\bibinfo {volume} {509}},\ \bibinfo
  {pages} {448} (\bibinfo {year} {1998})}\BibitemShut {NoStop}%
\bibitem [{\citenamefont {{de Arcangelis}}\ \emph {et~al.}(2006)\citenamefont
  {{de Arcangelis}}, \citenamefont {Godano}, \citenamefont {Lippiello},\ and\
  \citenamefont {Nicodemi}}]{deArcangelis2006Universality}%
  \BibitemOpen
  \bibfield  {author} {\bibinfo {author} {\bibfnamefont {L.}~\bibnamefont {{de
  Arcangelis}}}, \bibinfo {author} {\bibfnamefont {C.}~\bibnamefont {Godano}},
  \bibinfo {author} {\bibfnamefont {E.}~\bibnamefont {Lippiello}},\ and\
  \bibinfo {author} {\bibfnamefont {M.}~\bibnamefont {Nicodemi}},\ }\bibfield
  {title} {\bibinfo {title} {Universality in solar flare and earthquake
  occurrence},\ }\href {https://doi.org/10.1103/physrevlett.96.051102}
  {\bibfield  {journal} {\bibinfo  {journal} {Physical Review Letters}\
  }\textbf {\bibinfo {volume} {96}},\ \bibinfo {pages} {051102} (\bibinfo
  {year} {2006})}\BibitemShut {NoStop}%
\bibitem [{\citenamefont {Bak}\ \emph {et~al.}(2002)\citenamefont {Bak},
  \citenamefont {Christensen}, \citenamefont {Danon},\ and\ \citenamefont
  {Scanlon}}]{Bak2002Unified}%
  \BibitemOpen
  \bibfield  {author} {\bibinfo {author} {\bibfnamefont {P.}~\bibnamefont
  {Bak}}, \bibinfo {author} {\bibfnamefont {K.}~\bibnamefont {Christensen}},
  \bibinfo {author} {\bibfnamefont {L.}~\bibnamefont {Danon}},\ and\ \bibinfo
  {author} {\bibfnamefont {T.}~\bibnamefont {Scanlon}},\ }\bibfield  {title}
  {\bibinfo {title} {Unified scaling law for earthquakes},\ }\href
  {https://doi.org/10.1103/physrevlett.88.178501} {\bibfield  {journal}
  {\bibinfo  {journal} {Physical Review Letters}\ }\textbf {\bibinfo {volume}
  {88}},\ \bibinfo {pages} {178501} (\bibinfo {year} {2002})}\BibitemShut
  {NoStop}%
\bibitem [{\citenamefont {Corral}(2004)}]{Corral2004Longterm}%
  \BibitemOpen
  \bibfield  {author} {\bibinfo {author} {\bibfnamefont {{\'A}.}~\bibnamefont
  {Corral}},\ }\bibfield  {title} {\bibinfo {title} {Long-term clustering,
  scaling, and universality in the temporal occurrence of earthquakes},\ }\href
  {https://doi.org/10.1103/physrevlett.92.108501} {\bibfield  {journal}
  {\bibinfo  {journal} {Physical Review Letters}\ }\textbf {\bibinfo {volume}
  {92}},\ \bibinfo {pages} {108501} (\bibinfo {year} {2004})}\BibitemShut
  {NoStop}%
\bibitem [{\citenamefont {Beggs}\ and\ \citenamefont
  {Plenz}(2003)}]{Beggs2003Neuronal}%
  \BibitemOpen
  \bibfield  {author} {\bibinfo {author} {\bibfnamefont {J.~M.}\ \bibnamefont
  {Beggs}}\ and\ \bibinfo {author} {\bibfnamefont {D.}~\bibnamefont {Plenz}},\
  }\bibfield  {title} {\bibinfo {title} {Neuronal {{Avalanches}} in
  {{Neocortical Circuits}}},\ }\href
  {https://doi.org/10.1523/JNEUROSCI.23-35-11167.2003} {\bibfield  {journal}
  {\bibinfo  {journal} {The Journal of Neuroscience}\ }\textbf {\bibinfo
  {volume} {23}},\ \bibinfo {pages} {11167} (\bibinfo {year}
  {2003})}\BibitemShut {NoStop}%
\bibitem [{\citenamefont {Petermann}\ \emph {et~al.}(2009)\citenamefont
  {Petermann}, \citenamefont {Thiagarajan}, \citenamefont {Lebedev},
  \citenamefont {Nicolelis}, \citenamefont {Chialvo},\ and\ \citenamefont
  {Plenz}}]{Petermann2009Spontaneous}%
  \BibitemOpen
  \bibfield  {author} {\bibinfo {author} {\bibfnamefont {T.}~\bibnamefont
  {Petermann}}, \bibinfo {author} {\bibfnamefont {T.~C.}\ \bibnamefont
  {Thiagarajan}}, \bibinfo {author} {\bibfnamefont {M.~A.}\ \bibnamefont
  {Lebedev}}, \bibinfo {author} {\bibfnamefont {M.~A.~L.}\ \bibnamefont
  {Nicolelis}}, \bibinfo {author} {\bibfnamefont {D.~R.}\ \bibnamefont
  {Chialvo}},\ and\ \bibinfo {author} {\bibfnamefont {D.}~\bibnamefont
  {Plenz}},\ }\bibfield  {title} {\bibinfo {title} {Spontaneous cortical
  activity in awake monkeys composed of neuronal avalanches},\ }\href
  {https://doi.org/10.1073/pnas.0904089106} {\bibfield  {journal} {\bibinfo
  {journal} {Proceedings of the National Academy of Sciences}\ }\textbf
  {\bibinfo {volume} {106}},\ \bibinfo {pages} {15921} (\bibinfo {year}
  {2009})}\BibitemShut {NoStop}%
\bibitem [{\citenamefont {Kemuriyama}\ \emph {et~al.}(2010)\citenamefont
  {Kemuriyama}, \citenamefont {Ohta}, \citenamefont {Sato}, \citenamefont
  {Maruyama}, \citenamefont {{Tandai-Hiruma}}, \citenamefont {Kato},\ and\
  \citenamefont {Nishida}}]{Kemuriyama2010Powerlaw}%
  \BibitemOpen
  \bibfield  {author} {\bibinfo {author} {\bibfnamefont {T.}~\bibnamefont
  {Kemuriyama}}, \bibinfo {author} {\bibfnamefont {H.}~\bibnamefont {Ohta}},
  \bibinfo {author} {\bibfnamefont {Y.}~\bibnamefont {Sato}}, \bibinfo {author}
  {\bibfnamefont {S.}~\bibnamefont {Maruyama}}, \bibinfo {author}
  {\bibfnamefont {M.}~\bibnamefont {{Tandai-Hiruma}}}, \bibinfo {author}
  {\bibfnamefont {K.}~\bibnamefont {Kato}},\ and\ \bibinfo {author}
  {\bibfnamefont {Y.}~\bibnamefont {Nishida}},\ }\bibfield  {title} {\bibinfo
  {title} {A power-law distribution of inter-spike intervals in renal
  sympathetic nerve activity in salt-sensitive hypertension-induced chronic
  heart failure},\ }\href {https://doi.org/10.1016/j.biosystems.2010.06.002}
  {\bibfield  {journal} {\bibinfo  {journal} {BioSystems}\ }\textbf {\bibinfo
  {volume} {101}},\ \bibinfo {pages} {144} (\bibinfo {year}
  {2010})}\BibitemShut {NoStop}%
\bibitem [{\citenamefont {Harder}\ and\ \citenamefont
  {Paczuski}(2006)}]{Harder2006Correlated}%
  \BibitemOpen
  \bibfield  {author} {\bibinfo {author} {\bibfnamefont {U.}~\bibnamefont
  {Harder}}\ and\ \bibinfo {author} {\bibfnamefont {M.}~\bibnamefont
  {Paczuski}},\ }\bibfield  {title} {\bibinfo {title} {Correlated dynamics in
  human printing behavior},\ }\href
  {https://doi.org/10.1016/j.physa.2005.06.079} {\bibfield  {journal} {\bibinfo
   {journal} {Physica A: Statistical Mechanics and its Applications}\ }\textbf
  {\bibinfo {volume} {361}},\ \bibinfo {pages} {329} (\bibinfo {year}
  {2006})}\BibitemShut {NoStop}%
\bibitem [{\citenamefont {V{\'a}zquez}\ \emph {et~al.}(2006)\citenamefont
  {V{\'a}zquez}, \citenamefont {Oliveira}, \citenamefont {Dezs{\"o}},
  \citenamefont {Goh}, \citenamefont {Kondor},\ and\ \citenamefont
  {Barab{\'a}si}}]{Vazquez2006Modeling}%
  \BibitemOpen
  \bibfield  {author} {\bibinfo {author} {\bibfnamefont {A.}~\bibnamefont
  {V{\'a}zquez}}, \bibinfo {author} {\bibfnamefont {J.~G.}\ \bibnamefont
  {Oliveira}}, \bibinfo {author} {\bibfnamefont {Z.}~\bibnamefont {Dezs{\"o}}},
  \bibinfo {author} {\bibfnamefont {K.-I.}\ \bibnamefont {Goh}}, \bibinfo
  {author} {\bibfnamefont {I.}~\bibnamefont {Kondor}},\ and\ \bibinfo {author}
  {\bibfnamefont {A.-L.}\ \bibnamefont {Barab{\'a}si}},\ }\bibfield  {title}
  {\bibinfo {title} {Modeling bursts and heavy tails in human dynamics},\
  }\href {https://doi.org/10.1103/physreve.73.036127} {\bibfield  {journal}
  {\bibinfo  {journal} {Physical Review E}\ }\textbf {\bibinfo {volume} {73}},\
  \bibinfo {pages} {036127} (\bibinfo {year} {2006})}\BibitemShut {NoStop}%
\bibitem [{\citenamefont {Goh}\ and\ \citenamefont
  {Barab{\'a}si}(2008)}]{Goh2008Burstiness}%
  \BibitemOpen
  \bibfield  {author} {\bibinfo {author} {\bibfnamefont {K.-I.}\ \bibnamefont
  {Goh}}\ and\ \bibinfo {author} {\bibfnamefont {A.-L.}\ \bibnamefont
  {Barab{\'a}si}},\ }\bibfield  {title} {\bibinfo {title} {Burstiness and
  memory in complex systems},\ }\href
  {https://doi.org/10.1209/0295-5075/81/48002} {\bibfield  {journal} {\bibinfo
  {journal} {EPL (Europhysics Letters)}\ }\textbf {\bibinfo {volume} {81}},\
  \bibinfo {pages} {48002} (\bibinfo {year} {2008})}\BibitemShut {NoStop}%
\bibitem [{\citenamefont {Rybski}\ \emph {et~al.}(2009)\citenamefont {Rybski},
  \citenamefont {Buldyrev}, \citenamefont {Havlin}, \citenamefont {Liljeros},\
  and\ \citenamefont {Makse}}]{Rybski2009Scaling}%
  \BibitemOpen
  \bibfield  {author} {\bibinfo {author} {\bibfnamefont {D.}~\bibnamefont
  {Rybski}}, \bibinfo {author} {\bibfnamefont {S.~V.}\ \bibnamefont
  {Buldyrev}}, \bibinfo {author} {\bibfnamefont {S.}~\bibnamefont {Havlin}},
  \bibinfo {author} {\bibfnamefont {F.}~\bibnamefont {Liljeros}},\ and\
  \bibinfo {author} {\bibfnamefont {H.~A.}\ \bibnamefont {Makse}},\ }\bibfield
  {title} {\bibinfo {title} {Scaling laws of human interaction activity},\
  }\href {https://doi.org/10.1073/pnas.0902667106} {\bibfield  {journal}
  {\bibinfo  {journal} {Proceedings of the National Academy of Sciences}\
  }\textbf {\bibinfo {volume} {106}},\ \bibinfo {pages} {12640} (\bibinfo
  {year} {2009})}\BibitemShut {NoStop}%
\bibitem [{\citenamefont {Wu}\ \emph {et~al.}(2010)\citenamefont {Wu},
  \citenamefont {Zhou}, \citenamefont {Xiao}, \citenamefont {Kurths},\ and\
  \citenamefont {Schellnhuber}}]{Wu2010Evidence}%
  \BibitemOpen
  \bibfield  {author} {\bibinfo {author} {\bibfnamefont {Y.}~\bibnamefont
  {Wu}}, \bibinfo {author} {\bibfnamefont {C.}~\bibnamefont {Zhou}}, \bibinfo
  {author} {\bibfnamefont {J.}~\bibnamefont {Xiao}}, \bibinfo {author}
  {\bibfnamefont {J.}~\bibnamefont {Kurths}},\ and\ \bibinfo {author}
  {\bibfnamefont {H.~J.}\ \bibnamefont {Schellnhuber}},\ }\bibfield  {title}
  {\bibinfo {title} {Evidence for a bimodal distribution in human
  communication.},\ }\href {https://doi.org/10.1073/pnas.1013140107} {\bibfield
   {journal} {\bibinfo  {journal} {Proceedings of the National Academy of
  Sciences}\ }\textbf {\bibinfo {volume} {107}},\ \bibinfo {pages} {18803}
  (\bibinfo {year} {2010})}\BibitemShut {NoStop}%
\bibitem [{\citenamefont {Jiang}\ \emph {et~al.}(2013)\citenamefont {Jiang},
  \citenamefont {Xie}, \citenamefont {Li}, \citenamefont {Podobnik},
  \citenamefont {Zhou},\ and\ \citenamefont {Stanley}}]{Jiang2013Calling}%
  \BibitemOpen
  \bibfield  {author} {\bibinfo {author} {\bibfnamefont {Z.-Q.}\ \bibnamefont
  {Jiang}}, \bibinfo {author} {\bibfnamefont {W.-J.}\ \bibnamefont {Xie}},
  \bibinfo {author} {\bibfnamefont {M.-X.}\ \bibnamefont {Li}}, \bibinfo
  {author} {\bibfnamefont {B.}~\bibnamefont {Podobnik}}, \bibinfo {author}
  {\bibfnamefont {W.-X.}\ \bibnamefont {Zhou}},\ and\ \bibinfo {author}
  {\bibfnamefont {H.~E.}\ \bibnamefont {Stanley}},\ }\bibfield  {title}
  {\bibinfo {title} {Calling patterns in human communication dynamics},\ }\href
  {https://doi.org/10.1073/pnas.1220433110} {\bibfield  {journal} {\bibinfo
  {journal} {Proceedings of the National Academy of Sciences}\ }\textbf
  {\bibinfo {volume} {110}},\ \bibinfo {pages} {1600} (\bibinfo {year}
  {2013})}\BibitemShut {NoStop}%
\bibitem [{\citenamefont {Fournet}\ and\ \citenamefont
  {Barrat}(2014)}]{Fournet2014Contact}%
  \BibitemOpen
  \bibfield  {author} {\bibinfo {author} {\bibfnamefont {J.}~\bibnamefont
  {Fournet}}\ and\ \bibinfo {author} {\bibfnamefont {A.}~\bibnamefont
  {Barrat}},\ }\bibfield  {title} {\bibinfo {title} {Contact {{Patterns}} among
  {{High School Students}}},\ }\href
  {https://doi.org/10.1371/journal.pone.0107878} {\bibfield  {journal}
  {\bibinfo  {journal} {PLoS ONE}\ }\textbf {\bibinfo {volume} {9}},\ \bibinfo
  {pages} {e107878} (\bibinfo {year} {2014})}\BibitemShut {NoStop}%
\bibitem [{\citenamefont {Jo}\ \emph {et~al.}(2020)\citenamefont {Jo},
  \citenamefont {Hiraoka},\ and\ \citenamefont {Kivel{\"a}}}]{Jo2020Bursttree}%
  \BibitemOpen
  \bibfield  {author} {\bibinfo {author} {\bibfnamefont {H.-H.}\ \bibnamefont
  {Jo}}, \bibinfo {author} {\bibfnamefont {T.}~\bibnamefont {Hiraoka}},\ and\
  \bibinfo {author} {\bibfnamefont {M.}~\bibnamefont {Kivel{\"a}}},\ }\bibfield
   {title} {\bibinfo {title} {Burst-tree decomposition of time series reveals
  the structure of temporal correlations},\ }\href
  {https://doi.org/10.1038/s41598-020-68157-1} {\bibfield  {journal} {\bibinfo
  {journal} {Scientific Reports}\ }\textbf {\bibinfo {volume} {10}},\ \bibinfo
  {pages} {12202} (\bibinfo {year} {2020})}\BibitemShut {NoStop}%
\bibitem [{\citenamefont {Choi}\ \emph {et~al.}(2021)\citenamefont {Choi},
  \citenamefont {Hiraoka},\ and\ \citenamefont
  {Jo}}]{Choi2021Individualdriven}%
  \BibitemOpen
  \bibfield  {author} {\bibinfo {author} {\bibfnamefont {J.}~\bibnamefont
  {Choi}}, \bibinfo {author} {\bibfnamefont {T.}~\bibnamefont {Hiraoka}},\ and\
  \bibinfo {author} {\bibfnamefont {H.-H.}\ \bibnamefont {Jo}},\ }\bibfield
  {title} {\bibinfo {title} {Individual-driven versus interaction-driven
  burstiness in human dynamics: {{The}} case of {{Wikipedia}} edit history},\
  }\href {https://doi.org/10.1103/PhysRevE.104.014312} {\bibfield  {journal}
  {\bibinfo  {journal} {Physical Review E}\ }\textbf {\bibinfo {volume}
  {104}},\ \bibinfo {pages} {014312} (\bibinfo {year} {2021})}\BibitemShut
  {NoStop}%
\bibitem [{\citenamefont {Bak}\ \emph {et~al.}(1987)\citenamefont {Bak},
  \citenamefont {Tang},\ and\ \citenamefont
  {Wiesenfeld}}]{Bak1987Selforganized}%
  \BibitemOpen
  \bibfield  {author} {\bibinfo {author} {\bibfnamefont {P.}~\bibnamefont
  {Bak}}, \bibinfo {author} {\bibfnamefont {C.}~\bibnamefont {Tang}},\ and\
  \bibinfo {author} {\bibfnamefont {K.}~\bibnamefont {Wiesenfeld}},\ }\bibfield
   {title} {\bibinfo {title} {Self-organized criticality: {{An}} explanation of
  1/f noise},\ }\href {https://doi.org/10.1103/physrevlett.59.381} {\bibfield
  {journal} {\bibinfo  {journal} {Physical Review Letters}\ }\textbf {\bibinfo
  {volume} {59}},\ \bibinfo {pages} {381} (\bibinfo {year} {1987})}\BibitemShut
  {NoStop}%
\bibitem [{\citenamefont {Jensen}(1998)}]{Jensen1998Selforganized}%
  \BibitemOpen
  \bibfield  {author} {\bibinfo {author} {\bibfnamefont {H.~J.}\ \bibnamefont
  {Jensen}},\ }\href {https://doi.org/10.1017/CBO9780511622717} {\emph
  {\bibinfo {title} {Self-{{Organized Criticality}}: {{Emergent Complex
  Behavior}} in {{Physical}} and {{Biological Systems}}}}},\ \bibinfo {edition}
  {1st}\ ed.\ (\bibinfo  {publisher} {Cambridge University Press},\ \bibinfo
  {year} {1998})\BibitemShut {NoStop}%
\bibitem [{\citenamefont {Christensen}\ and\ \citenamefont
  {Moloney}(2005)}]{Christensen2005Complexity}%
  \BibitemOpen
  \bibfield  {author} {\bibinfo {author} {\bibfnamefont {K.}~\bibnamefont
  {Christensen}}\ and\ \bibinfo {author} {\bibfnamefont {N.~R.}\ \bibnamefont
  {Moloney}},\ }\href@noop {} {\emph {\bibinfo {title} {Complexity and
  Criticality}}}\ (\bibinfo  {publisher} {Imperial College Press},\ \bibinfo
  {year} {2005})\BibitemShut {NoStop}%
\bibitem [{\citenamefont {Karsai}\ \emph {et~al.}(2012)\citenamefont {Karsai},
  \citenamefont {Kaski}, \citenamefont {Barab{\'a}si},\ and\ \citenamefont
  {Kert{\'e}sz}}]{Karsai2012Universal}%
  \BibitemOpen
  \bibfield  {author} {\bibinfo {author} {\bibfnamefont {M.}~\bibnamefont
  {Karsai}}, \bibinfo {author} {\bibfnamefont {K.}~\bibnamefont {Kaski}},
  \bibinfo {author} {\bibfnamefont {A.-L.}\ \bibnamefont {Barab{\'a}si}},\ and\
  \bibinfo {author} {\bibfnamefont {J.}~\bibnamefont {Kert{\'e}sz}},\
  }\bibfield  {title} {\bibinfo {title} {Universal features of correlated
  bursty behaviour},\ }\href {https://doi.org/10.1038/srep00397} {\bibfield
  {journal} {\bibinfo  {journal} {Scientific Reports}\ }\textbf {\bibinfo
  {volume} {2}},\ \bibinfo {pages} {397} (\bibinfo {year} {2012})}\BibitemShut
  {NoStop}%
\bibitem [{\citenamefont {Masuda}\ and\ \citenamefont
  {Rocha}(2018)}]{Masuda2018Gillespie}%
  \BibitemOpen
  \bibfield  {author} {\bibinfo {author} {\bibfnamefont {N.}~\bibnamefont
  {Masuda}}\ and\ \bibinfo {author} {\bibfnamefont {L.~E.~C.}\ \bibnamefont
  {Rocha}},\ }\bibfield  {title} {\bibinfo {title} {A {{Gillespie}} algorithm
  for non-{{Markovian}} stochastic processes},\ }\href
  {https://doi.org/10.1137/16m1055876} {\bibfield  {journal} {\bibinfo
  {journal} {SIAM Review}\ }\textbf {\bibinfo {volume} {60}},\ \bibinfo {pages}
  {95} (\bibinfo {year} {2018})}\BibitemShut {NoStop}%
\bibitem [{\citenamefont {Hiraoka}\ and\ \citenamefont
  {Jo}(2025)}]{Hiraoka2025Hierarchical}%
  \BibitemOpen
  \bibfield  {author} {\bibinfo {author} {\bibfnamefont {T.}~\bibnamefont
  {Hiraoka}}\ and\ \bibinfo {author} {\bibfnamefont {H.-H.}\ \bibnamefont
  {Jo}},\ }\href {https://doi.org/10.48550/arXiv.2508.18281} {\bibinfo {title}
  {Hierarchical organization of bursty trains in event sequences}} (\bibinfo
  {year} {2025}),\ \Eprint {https://arxiv.org/abs/2508.18281} {arXiv:2508.18281
  [physics]} \BibitemShut {NoStop}%
\bibitem [{\citenamefont {Birhanu}\ and\ \citenamefont
  {Jo}(2025{\natexlab{a}})}]{Birhanu2025Bursttree}%
  \BibitemOpen
  \bibfield  {author} {\bibinfo {author} {\bibfnamefont {T.}~\bibnamefont
  {Birhanu}}\ and\ \bibinfo {author} {\bibfnamefont {H.-H.}\ \bibnamefont
  {Jo}},\ }\bibfield  {title} {\bibinfo {title} {Burst-tree structure and
  higher-order temporal correlations},\ }\href
  {https://doi.org/10.1103/PhysRevE.111.014308} {\bibfield  {journal} {\bibinfo
   {journal} {Physical Review E}\ }\textbf {\bibinfo {volume} {111}},\ \bibinfo
  {pages} {014308} (\bibinfo {year} {2025}{\natexlab{a}})}\BibitemShut
  {NoStop}%
\bibitem [{\citenamefont {Birhanu}\ and\ \citenamefont
  {Jo}(2025{\natexlab{b}})}]{Birhanu2025Maximum}%
  \BibitemOpen
  \bibfield  {author} {\bibinfo {author} {\bibfnamefont {T.}~\bibnamefont
  {Birhanu}}\ and\ \bibinfo {author} {\bibfnamefont {H.-H.}\ \bibnamefont
  {Jo}},\ }\bibfield  {title} {\bibinfo {title} {Maximum likelihood estimation
  of burst-merging kernels for bursty time series},\ }\href
  {https://doi.org/10.1103/PhysRevE.111.054317} {\bibfield  {journal} {\bibinfo
   {journal} {Physical Review E}\ }\textbf {\bibinfo {volume} {111}},\ \bibinfo
  {pages} {054317} (\bibinfo {year} {2025}{\natexlab{b}})}\BibitemShut
  {NoStop}%
\bibitem [{\citenamefont {Peng}\ \emph {et~al.}(1994)\citenamefont {Peng},
  \citenamefont {Buldyrev}, \citenamefont {Havlin}, \citenamefont {Simons},
  \citenamefont {Stanley},\ and\ \citenamefont {Goldberger}}]{Peng1994Mosaic}%
  \BibitemOpen
  \bibfield  {author} {\bibinfo {author} {\bibfnamefont {C.-K.}\ \bibnamefont
  {Peng}}, \bibinfo {author} {\bibfnamefont {S.~V.}\ \bibnamefont {Buldyrev}},
  \bibinfo {author} {\bibfnamefont {S.}~\bibnamefont {Havlin}}, \bibinfo
  {author} {\bibfnamefont {M.}~\bibnamefont {Simons}}, \bibinfo {author}
  {\bibfnamefont {H.~E.}\ \bibnamefont {Stanley}},\ and\ \bibinfo {author}
  {\bibfnamefont {A.~L.}\ \bibnamefont {Goldberger}},\ }\bibfield  {title}
  {\bibinfo {title} {Mosaic organization of {{DNA}} nucleotides},\ }\href
  {https://doi.org/10.1103/PhysRevE.49.1685} {\bibfield  {journal} {\bibinfo
  {journal} {Physical Review E}\ }\textbf {\bibinfo {volume} {49}},\ \bibinfo
  {pages} {1685} (\bibinfo {year} {1994})}\BibitemShut {NoStop}%
\bibitem [{\citenamefont {Peng}\ \emph {et~al.}(1995)\citenamefont {Peng},
  \citenamefont {Havlin}, \citenamefont {Stanley},\ and\ \citenamefont
  {Goldberger}}]{Peng1995Quantification}%
  \BibitemOpen
  \bibfield  {author} {\bibinfo {author} {\bibfnamefont {C.-K.}\ \bibnamefont
  {Peng}}, \bibinfo {author} {\bibfnamefont {S.}~\bibnamefont {Havlin}},
  \bibinfo {author} {\bibfnamefont {H.~E.}\ \bibnamefont {Stanley}},\ and\
  \bibinfo {author} {\bibfnamefont {A.~L.}\ \bibnamefont {Goldberger}},\
  }\bibfield  {title} {\bibinfo {title} {Quantification of scaling exponents
  and crossover phenomena in nonstationary heartbeat time series},\ }\href
  {https://doi.org/10.1063/1.166141} {\bibfield  {journal} {\bibinfo  {journal}
  {Chaos: An Interdisciplinary Journal of Nonlinear Science}\ }\textbf
  {\bibinfo {volume} {5}},\ \bibinfo {pages} {82} (\bibinfo {year}
  {1995})}\BibitemShut {NoStop}%
\bibitem [{\citenamefont {Molkkari}\ \emph {et~al.}(2020)\citenamefont
  {Molkkari}, \citenamefont {Angelotti}, \citenamefont {Emig},\ and\
  \citenamefont {R{\"a}s{\"a}nen}}]{Molkkari2020Dynamical}%
  \BibitemOpen
  \bibfield  {author} {\bibinfo {author} {\bibfnamefont {M.}~\bibnamefont
  {Molkkari}}, \bibinfo {author} {\bibfnamefont {G.}~\bibnamefont {Angelotti}},
  \bibinfo {author} {\bibfnamefont {T.}~\bibnamefont {Emig}},\ and\ \bibinfo
  {author} {\bibfnamefont {E.}~\bibnamefont {R{\"a}s{\"a}nen}},\ }\bibfield
  {title} {\bibinfo {title} {Dynamical heart beat correlations during
  running},\ }\href {https://doi.org/10.1038/s41598-020-70358-7} {\bibfield
  {journal} {\bibinfo  {journal} {Scientific Reports}\ }\textbf {\bibinfo
  {volume} {10}},\ \bibinfo {pages} {13627} (\bibinfo {year}
  {2020})}\BibitemShut {NoStop}%
\bibitem [{\citenamefont {Pukkila}\ \emph {et~al.}(2025)\citenamefont
  {Pukkila}, \citenamefont {Molkkari}, \citenamefont {Hernesniemi},
  \citenamefont {Kanniainen},\ and\ \citenamefont
  {R{\"a}s{\"a}nen}}]{Pukkila2025Detection}%
  \BibitemOpen
  \bibfield  {author} {\bibinfo {author} {\bibfnamefont {T.}~\bibnamefont
  {Pukkila}}, \bibinfo {author} {\bibfnamefont {M.}~\bibnamefont {Molkkari}},
  \bibinfo {author} {\bibfnamefont {J.}~\bibnamefont {Hernesniemi}}, \bibinfo
  {author} {\bibfnamefont {M.}~\bibnamefont {Kanniainen}},\ and\ \bibinfo
  {author} {\bibfnamefont {E.}~\bibnamefont {R{\"a}s{\"a}nen}},\ }\bibfield
  {title} {\bibinfo {title} {Detection of congestive heart failure from {{RR}}
  intervals during long-term electrocardiographic recordings},\ }\href
  {https://doi.org/10.1016/j.hroo.2025.01.014} {\bibfield  {journal} {\bibinfo
  {journal} {Heart Rhythm O2}\ }\textbf {\bibinfo {volume} {6}},\ \bibinfo
  {pages} {509} (\bibinfo {year} {2025})}\BibitemShut {NoStop}%
\bibitem [{\citenamefont {Ramesh}\ \emph {et~al.}(2021)\citenamefont {Ramesh},
  \citenamefont {Solatidehkordi}, \citenamefont {Aburukba},\ and\ \citenamefont
  {Sagahyroon}}]{Ramesh2021Atrial}%
  \BibitemOpen
  \bibfield  {author} {\bibinfo {author} {\bibfnamefont {J.}~\bibnamefont
  {Ramesh}}, \bibinfo {author} {\bibfnamefont {Z.}~\bibnamefont
  {Solatidehkordi}}, \bibinfo {author} {\bibfnamefont {R.}~\bibnamefont
  {Aburukba}},\ and\ \bibinfo {author} {\bibfnamefont {A.}~\bibnamefont
  {Sagahyroon}},\ }\bibfield  {title} {\bibinfo {title} {Atrial {{Fibrillation
  Classification}} with {{Smart Wearables Using Short-Term Heart Rate
  Variability}} and {{Deep Convolutional Neural Networks}}},\ }\href
  {https://doi.org/10.3390/s21217233} {\bibfield  {journal} {\bibinfo
  {journal} {Sensors}\ }\textbf {\bibinfo {volume} {21}},\ \bibinfo {pages}
  {7233} (\bibinfo {year} {2021})}\BibitemShut {NoStop}%
\bibitem [{\citenamefont {{\c C}{\i}nar}\ and\ \citenamefont
  {Tuncer}(2021)}]{Cinar2021Classification}%
  \BibitemOpen
  \bibfield  {author} {\bibinfo {author} {\bibfnamefont {A.}~\bibnamefont {{\c
  C}{\i}nar}}\ and\ \bibinfo {author} {\bibfnamefont {S.~A.}\ \bibnamefont
  {Tuncer}},\ }\bibfield  {title} {\bibinfo {title} {Classification of normal
  sinus rhythm, abnormal arrhythmia and congestive heart failure {{ECG}}
  signals using {{LSTM}} and hybrid {{CNN-SVM}} deep neural networks},\ }\href
  {https://doi.org/10.1080/10255842.2020.1821192} {\bibfield  {journal}
  {\bibinfo  {journal} {Computer Methods in Biomechanics and Biomedical
  Engineering}\ }\textbf {\bibinfo {volume} {24}},\ \bibinfo {pages} {203}
  (\bibinfo {year} {2021})}\BibitemShut {NoStop}%
\bibitem [{\citenamefont {Richman}\ and\ \citenamefont
  {Moorman}(2000)}]{Richman2000Physiological}%
  \BibitemOpen
  \bibfield  {author} {\bibinfo {author} {\bibfnamefont {J.~S.}\ \bibnamefont
  {Richman}}\ and\ \bibinfo {author} {\bibfnamefont {J.~R.}\ \bibnamefont
  {Moorman}},\ }\bibfield  {title} {\bibinfo {title} {Physiological time-series
  analysis using approximate entropy and sample entropy},\ }\href
  {https://doi.org/10.1152/ajpheart.2000.278.6.H2039} {\bibfield  {journal}
  {\bibinfo  {journal} {American Journal of Physiology-Heart and Circulatory
  Physiology}\ }\textbf {\bibinfo {volume} {278}},\ \bibinfo {pages} {H2039}
  (\bibinfo {year} {2000})}\BibitemShut {NoStop}%
\bibitem [{\citenamefont {Costa}\ \emph {et~al.}(2002)\citenamefont {Costa},
  \citenamefont {Goldberger},\ and\ \citenamefont
  {Peng}}]{Costa2002Multiscale}%
  \BibitemOpen
  \bibfield  {author} {\bibinfo {author} {\bibfnamefont {M.}~\bibnamefont
  {Costa}}, \bibinfo {author} {\bibfnamefont {A.~L.}\ \bibnamefont
  {Goldberger}},\ and\ \bibinfo {author} {\bibfnamefont {C.-K.}\ \bibnamefont
  {Peng}},\ }\bibfield  {title} {\bibinfo {title} {Multiscale {{Entropy
  Analysis}} of {{Complex Physiologic Time Series}}},\ }\href
  {https://doi.org/10.1103/PhysRevLett.89.068102} {\bibfield  {journal}
  {\bibinfo  {journal} {Physical Review Letters}\ }\textbf {\bibinfo {volume}
  {89}},\ \bibinfo {pages} {068102} (\bibinfo {year} {2002})}\BibitemShut
  {NoStop}%
\bibitem [{\citenamefont {Costa}\ and\ \citenamefont
  {Goldberger}(2015)}]{Costa2015Generalized}%
  \BibitemOpen
  \bibfield  {author} {\bibinfo {author} {\bibfnamefont {M.}~\bibnamefont
  {Costa}}\ and\ \bibinfo {author} {\bibfnamefont {A.}~\bibnamefont
  {Goldberger}},\ }\bibfield  {title} {\bibinfo {title} {Generalized
  {{Multiscale Entropy Analysis}}: {{Application}} to {{Quantifying}} the
  {{Complex Volatility}} of {{Human Heartbeat Time Series}}},\ }\href
  {https://doi.org/10.3390/e17031197} {\bibfield  {journal} {\bibinfo
  {journal} {Entropy}\ }\textbf {\bibinfo {volume} {17}},\ \bibinfo {pages}
  {1197} (\bibinfo {year} {2015})}\BibitemShut {NoStop}%
\bibitem [{\citenamefont {{Mu{\~n}oz-Diosdado}}\ \emph
  {et~al.}(2023)\citenamefont {{Mu{\~n}oz-Diosdado}}, \citenamefont
  {{Sol{\'i}s-Montufar}},\ and\ \citenamefont
  {{Zamora-Justo}}}]{Munoz-Diosdado2023Visibility}%
  \BibitemOpen
  \bibfield  {author} {\bibinfo {author} {\bibfnamefont {A.}~\bibnamefont
  {{Mu{\~n}oz-Diosdado}}}, \bibinfo {author} {\bibfnamefont {{\'E}.~E.}\
  \bibnamefont {{Sol{\'i}s-Montufar}}},\ and\ \bibinfo {author} {\bibfnamefont
  {J.~A.}\ \bibnamefont {{Zamora-Justo}}},\ }\bibfield  {title} {\bibinfo
  {title} {Visibility {{Graph Analysis}} of {{Heartbeat Time Series}}:
  {{Comparison}} of {{Young}} vs. {{Old}}, {{Healthy}} vs. {{Diseased}},
  {{Rest}} vs. {{Exercise}}, and {{Sedentary}} vs. {{Active}}},\ }\href
  {https://doi.org/10.3390/e25040677} {\bibfield  {journal} {\bibinfo
  {journal} {Entropy}\ }\textbf {\bibinfo {volume} {25}},\ \bibinfo {pages}
  {677} (\bibinfo {year} {2023})}\BibitemShut {NoStop}%
\bibitem [{Phy()}]{PhysioBank}%
  \BibitemOpen
  \href@noop {} {\bibinfo {title} {{{PhysioBank}}}},\ \bibinfo {howpublished}
  {https://physionet.org/physiobank/}\BibitemShut {NoStop}%
\bibitem [{\citenamefont {Kim}\ and\ \citenamefont
  {Jo}(2016)}]{Kim2016Measuring}%
  \BibitemOpen
  \bibfield  {author} {\bibinfo {author} {\bibfnamefont {E.-K.}\ \bibnamefont
  {Kim}}\ and\ \bibinfo {author} {\bibfnamefont {H.-H.}\ \bibnamefont {Jo}},\
  }\bibfield  {title} {\bibinfo {title} {Measuring burstiness for finite event
  sequences},\ }\href {https://doi.org/10.1103/physreve.94.032311} {\bibfield
  {journal} {\bibinfo  {journal} {Physical Review E}\ }\textbf {\bibinfo
  {volume} {94}},\ \bibinfo {pages} {032311} (\bibinfo {year}
  {2016})}\BibitemShut {NoStop}%
\bibitem [{\citenamefont {Jo}(2017)}]{Jo2017Modeling}%
  \BibitemOpen
  \bibfield  {author} {\bibinfo {author} {\bibfnamefont {H.-H.}\ \bibnamefont
  {Jo}},\ }\bibfield  {title} {\bibinfo {title} {Modeling correlated bursts by
  the bursty-get-burstier mechanism},\ }\href
  {https://doi.org/10.1103/physreve.96.062131} {\bibfield  {journal} {\bibinfo
  {journal} {Physical Review E}\ }\textbf {\bibinfo {volume} {96}},\ \bibinfo
  {pages} {062131} (\bibinfo {year} {2017})}\BibitemShut {NoStop}%
\bibitem [{\citenamefont {Barab{\'a}si}\ and\ \citenamefont
  {P{\'o}sfai}(2016)}]{Barabasi2016Network}%
  \BibitemOpen
  \bibfield  {author} {\bibinfo {author} {\bibfnamefont {A.-L.}\ \bibnamefont
  {Barab{\'a}si}}\ and\ \bibinfo {author} {\bibfnamefont {M.}~\bibnamefont
  {P{\'o}sfai}},\ }\href@noop {} {\emph {\bibinfo {title} {Network Science}}}\
  (\bibinfo  {publisher} {Cambridge University Press},\ \bibinfo {address}
  {Cambridge},\ \bibinfo {year} {2016})\BibitemShut {NoStop}%
\bibitem [{\citenamefont {Pham}\ \emph {et~al.}(2015)\citenamefont {Pham},
  \citenamefont {Sheridan},\ and\ \citenamefont {Shimodaira}}]{Pham2015PAFit}%
  \BibitemOpen
  \bibfield  {author} {\bibinfo {author} {\bibfnamefont {T.}~\bibnamefont
  {Pham}}, \bibinfo {author} {\bibfnamefont {P.}~\bibnamefont {Sheridan}},\
  and\ \bibinfo {author} {\bibfnamefont {H.}~\bibnamefont {Shimodaira}},\
  }\bibfield  {title} {\bibinfo {title} {{{PAFit}}: {{A Statistical Method}}
  for {{Measuring Preferential Attachment}} in {{Temporal Complex Networks}}},\
  }\href {https://doi.org/10.1371/journal.pone.0137796} {\bibfield  {journal}
  {\bibinfo  {journal} {PLoS ONE}\ }\textbf {\bibinfo {volume} {10}},\ \bibinfo
  {pages} {e0137796} (\bibinfo {year} {2015})}\BibitemShut {NoStop}%
\bibitem [{\citenamefont {Barab{\'a}si}\ and\ \citenamefont
  {Albert}(1999)}]{Barabasi1999Emergence}%
  \BibitemOpen
  \bibfield  {author} {\bibinfo {author} {\bibfnamefont {A.-L.}\ \bibnamefont
  {Barab{\'a}si}}\ and\ \bibinfo {author} {\bibfnamefont {R.}~\bibnamefont
  {Albert}},\ }\bibfield  {title} {\bibinfo {title} {Emergence of {{Scaling}}
  in {{Random Networks}}},\ }\href
  {https://doi.org/10.1126/science.286.5439.509} {\bibfield  {journal}
  {\bibinfo  {journal} {Science}\ }\textbf {\bibinfo {volume} {286}},\ \bibinfo
  {pages} {509} (\bibinfo {year} {1999})}\BibitemShut {NoStop}%
\bibitem [{\citenamefont {Krapivsky}\ \emph {et~al.}(2000)\citenamefont
  {Krapivsky}, \citenamefont {Redner},\ and\ \citenamefont
  {Leyvraz}}]{Krapivsky2000Connectivity}%
  \BibitemOpen
  \bibfield  {author} {\bibinfo {author} {\bibfnamefont {P.~L.}\ \bibnamefont
  {Krapivsky}}, \bibinfo {author} {\bibfnamefont {S.}~\bibnamefont {Redner}},\
  and\ \bibinfo {author} {\bibfnamefont {F.}~\bibnamefont {Leyvraz}},\
  }\bibfield  {title} {\bibinfo {title} {Connectivity of {{Growing Random
  Networks}}},\ }\href {https://doi.org/10.1103/PhysRevLett.85.4629} {\bibfield
   {journal} {\bibinfo  {journal} {Physical Review Letters}\ }\textbf {\bibinfo
  {volume} {85}},\ \bibinfo {pages} {4629} (\bibinfo {year}
  {2000})}\BibitemShut {NoStop}%
\bibitem [{\citenamefont {Newman}(2001)}]{Newman2001Clustering}%
  \BibitemOpen
  \bibfield  {author} {\bibinfo {author} {\bibfnamefont {M.~E.~J.}\
  \bibnamefont {Newman}},\ }\bibfield  {title} {\bibinfo {title} {Clustering
  and preferential attachment in growing networks},\ }\href
  {https://doi.org/10.1103/physreve.64.025102} {\bibfield  {journal} {\bibinfo
  {journal} {Physical Review E}\ }\textbf {\bibinfo {volume} {64}},\ \bibinfo
  {pages} {025102} (\bibinfo {year} {2001})}\BibitemShut {NoStop}%
\bibitem [{\citenamefont {Jeong}\ \emph {et~al.}(2003)\citenamefont {Jeong},
  \citenamefont {N{\'e}da},\ and\ \citenamefont
  {Barab{\'a}si}}]{Jeong2003Measuring}%
  \BibitemOpen
  \bibfield  {author} {\bibinfo {author} {\bibfnamefont {H.}~\bibnamefont
  {Jeong}}, \bibinfo {author} {\bibfnamefont {Z.}~\bibnamefont {N{\'e}da}},\
  and\ \bibinfo {author} {\bibfnamefont {A.~L.}\ \bibnamefont {Barab{\'a}si}},\
  }\bibfield  {title} {\bibinfo {title} {Measuring preferential attachment in
  evolving networks},\ }\href {https://doi.org/10.1209/epl/i2003-00166-9}
  {\bibfield  {journal} {\bibinfo  {journal} {Europhysics Letters (EPL)}\
  }\textbf {\bibinfo {volume} {61}},\ \bibinfo {pages} {567} (\bibinfo {year}
  {2003})}\BibitemShut {NoStop}%
\bibitem [{\citenamefont {Sheridan}\ \emph {et~al.}(2012)\citenamefont
  {Sheridan}, \citenamefont {Yagahara},\ and\ \citenamefont
  {Shimodaira}}]{Sheridan2012Measuring}%
  \BibitemOpen
  \bibfield  {author} {\bibinfo {author} {\bibfnamefont {P.}~\bibnamefont
  {Sheridan}}, \bibinfo {author} {\bibfnamefont {Y.}~\bibnamefont {Yagahara}},\
  and\ \bibinfo {author} {\bibfnamefont {H.}~\bibnamefont {Shimodaira}},\
  }\bibfield  {title} {\bibinfo {title} {Measuring preferential attachment in
  growing networks with missing-timelines using {{Markov}} chain {{Monte
  Carlo}}},\ }\href {https://doi.org/10.1016/j.physa.2012.05.041} {\bibfield
  {journal} {\bibinfo  {journal} {Physica A: Statistical Mechanics and its
  Applications}\ }\textbf {\bibinfo {volume} {391}},\ \bibinfo {pages} {5031}
  (\bibinfo {year} {2012})}\BibitemShut {NoStop}%
\bibitem [{\citenamefont {Kunegis}\ \emph {et~al.}(2013)\citenamefont
  {Kunegis}, \citenamefont {Blattner},\ and\ \citenamefont
  {Moser}}]{Kunegis2013Preferential}%
  \BibitemOpen
  \bibfield  {author} {\bibinfo {author} {\bibfnamefont {J.}~\bibnamefont
  {Kunegis}}, \bibinfo {author} {\bibfnamefont {M.}~\bibnamefont {Blattner}},\
  and\ \bibinfo {author} {\bibfnamefont {C.}~\bibnamefont {Moser}},\ }\bibfield
   {title} {\bibinfo {title} {Preferential attachment in online networks:
  Measurement and explanations},\ }in\ \href
  {https://doi.org/10.1145/2464464.2464514} {\emph {\bibinfo {booktitle}
  {Proceedings of the 5th {{Annual ACM Web Science Conference}}}}}\ (\bibinfo
  {publisher} {ACM},\ \bibinfo {address} {Paris France},\ \bibinfo {year}
  {2013})\ pp.\ \bibinfo {pages} {205--214}\BibitemShut {NoStop}%
\bibitem [{\citenamefont {Stockmayer}(1943)}]{Stockmayer1943Theory}%
  \BibitemOpen
  \bibfield  {author} {\bibinfo {author} {\bibfnamefont {W.~H.}\ \bibnamefont
  {Stockmayer}},\ }\bibfield  {title} {\bibinfo {title} {Theory of {{Molecular
  Size Distribution}} and {{Gel Formation}} in {{Branched-Chain Polymers}}},\
  }\href {https://doi.org/10.1063/1.1723803} {\bibfield  {journal} {\bibinfo
  {journal} {The Journal of Chemical Physics}\ }\textbf {\bibinfo {volume}
  {11}},\ \bibinfo {pages} {45} (\bibinfo {year} {1943})}\BibitemShut {NoStop}%
\bibitem [{\citenamefont {Lushnikov}(1973)}]{Lushnikov1973Evolution}%
  \BibitemOpen
  \bibfield  {author} {\bibinfo {author} {\bibfnamefont {A.}~\bibnamefont
  {Lushnikov}},\ }\bibfield  {title} {\bibinfo {title} {Evolution of
  coagulating systems},\ }\href {https://doi.org/10.1016/0021-9797(73)90171-9}
  {\bibfield  {journal} {\bibinfo  {journal} {Journal of Colloid and Interface
  Science}\ }\textbf {\bibinfo {volume} {45}},\ \bibinfo {pages} {549}
  (\bibinfo {year} {1973})}\BibitemShut {NoStop}%
\bibitem [{\citenamefont {White}(1982)}]{White1982Form}%
  \BibitemOpen
  \bibfield  {author} {\bibinfo {author} {\bibfnamefont {W.~H.}\ \bibnamefont
  {White}},\ }\bibfield  {title} {\bibinfo {title} {On the form of steady-state
  solutions to the coagulation equations},\ }\href
  {https://doi.org/10.1016/0021-9797(82)90382-4} {\bibfield  {journal}
  {\bibinfo  {journal} {Journal of Colloid and Interface Science}\ }\textbf
  {\bibinfo {volume} {87}},\ \bibinfo {pages} {204} (\bibinfo {year}
  {1982})}\BibitemShut {NoStop}%
\bibitem [{\citenamefont {Hendriks}\ \emph {et~al.}(1983)\citenamefont
  {Hendriks}, \citenamefont {Ernst},\ and\ \citenamefont
  {Ziff}}]{Hendriks1983Coagulation}%
  \BibitemOpen
  \bibfield  {author} {\bibinfo {author} {\bibfnamefont {E.~M.}\ \bibnamefont
  {Hendriks}}, \bibinfo {author} {\bibfnamefont {M.~H.}\ \bibnamefont
  {Ernst}},\ and\ \bibinfo {author} {\bibfnamefont {R.~M.}\ \bibnamefont
  {Ziff}},\ }\bibfield  {title} {\bibinfo {title} {Coagulation equations with
  gelation},\ }\href {https://doi.org/10.1007/BF01019497} {\bibfield  {journal}
  {\bibinfo  {journal} {Journal of Statistical Physics}\ }\textbf {\bibinfo
  {volume} {31}},\ \bibinfo {pages} {519} (\bibinfo {year} {1983})}\BibitemShut
  {NoStop}%
\bibitem [{\citenamefont {Aldous}(1999)}]{Aldous1999Deterministic}%
  \BibitemOpen
  \bibfield  {author} {\bibinfo {author} {\bibfnamefont {D.~J.}\ \bibnamefont
  {Aldous}},\ }\bibfield  {title} {\bibinfo {title} {Deterministic and
  {{Stochastic Models}} for {{Coalescence}} ({{Aggregation}} and
  {{Coagulation}}): {{A Review}} of the {{Mean-Field Theory}} for
  {{Probabilists}}},\ }\href {https://doi.org/10.2307/3318611} {\bibfield
  {journal} {\bibinfo  {journal} {Bernoulli}\ }\textbf {\bibinfo {volume}
  {5}},\ \bibinfo {pages} {3} (\bibinfo {year} {1999})}\BibitemShut {NoStop}%
\bibitem [{\citenamefont {Lee}(2001)}]{Lee2001Survey}%
  \BibitemOpen
  \bibfield  {author} {\bibinfo {author} {\bibfnamefont {M.~H.}\ \bibnamefont
  {Lee}},\ }\bibfield  {title} {\bibinfo {title} {A survey of numerical
  solutions to the coagulation equation},\ }\href
  {https://doi.org/10.1088/0305-4470/34/47/323} {\bibfield  {journal} {\bibinfo
   {journal} {Journal of Physics A: Mathematical and General}\ }\textbf
  {\bibinfo {volume} {34}},\ \bibinfo {pages} {10219} (\bibinfo {year}
  {2001})}\BibitemShut {NoStop}%
\bibitem [{\citenamefont {Leyvraz}(2003)}]{Leyvraz2003Scaling}%
  \BibitemOpen
  \bibfield  {author} {\bibinfo {author} {\bibfnamefont {F.}~\bibnamefont
  {Leyvraz}},\ }\bibfield  {title} {\bibinfo {title} {Scaling theory and
  exactly solved models in the kinetics of irreversible aggregation},\ }\href
  {https://doi.org/10.1016/S0370-1573(03)00241-2} {\bibfield  {journal}
  {\bibinfo  {journal} {Physics Reports}\ }\textbf {\bibinfo {volume} {383}},\
  \bibinfo {pages} {95} (\bibinfo {year} {2003})}\BibitemShut {NoStop}%
\bibitem [{\citenamefont {Leyvraz}(2005)}]{Leyvraz2005Rigorous}%
  \BibitemOpen
  \bibfield  {author} {\bibinfo {author} {\bibfnamefont {F.}~\bibnamefont
  {Leyvraz}},\ }\bibfield  {title} {\bibinfo {title} {Rigorous {{Results}} in
  the {{Scaling Theory}} of {{Irreversible Aggregation Kinetics}}:},\ }\href
  {https://doi.org/10.2991/jnmp.2005.12.s1.37} {\bibfield  {journal} {\bibinfo
  {journal} {Journal of Nonlinear Mathematical Physics}\ }\textbf {\bibinfo
  {volume} {12}},\ \bibinfo {pages} {449} (\bibinfo {year} {2005})}\BibitemShut
  {NoStop}%
\bibitem [{\citenamefont {Wattis}(2006)}]{Wattis2006Introduction}%
  \BibitemOpen
  \bibfield  {author} {\bibinfo {author} {\bibfnamefont {J.~A.}\ \bibnamefont
  {Wattis}},\ }\bibfield  {title} {\bibinfo {title} {An introduction to
  mathematical models of coagulation--fragmentation processes: {{A}} discrete
  deterministic mean-field approach},\ }\href
  {https://doi.org/10.1016/j.physd.2006.07.024} {\bibfield  {journal} {\bibinfo
   {journal} {Physica D: Nonlinear Phenomena}\ }\textbf {\bibinfo {volume}
  {222}},\ \bibinfo {pages} {1} (\bibinfo {year} {2006})}\BibitemShut {NoStop}%
\bibitem [{\citenamefont {Cortes}\ and\ \citenamefont
  {Vapnik}(1995)}]{Cortes1995Supportvector}%
  \BibitemOpen
  \bibfield  {author} {\bibinfo {author} {\bibfnamefont {C.}~\bibnamefont
  {Cortes}}\ and\ \bibinfo {author} {\bibfnamefont {V.}~\bibnamefont
  {Vapnik}},\ }\bibfield  {title} {\bibinfo {title} {Support-vector networks},\
  }\href {https://doi.org/10.1007/BF00994018} {\bibfield  {journal} {\bibinfo
  {journal} {Machine Learning}\ }\textbf {\bibinfo {volume} {20}},\ \bibinfo
  {pages} {273} (\bibinfo {year} {1995})}\BibitemShut {NoStop}%
\bibitem [{\citenamefont {Tharwat}(2021)}]{Tharwat2021Classification}%
  \BibitemOpen
  \bibfield  {author} {\bibinfo {author} {\bibfnamefont {A.}~\bibnamefont
  {Tharwat}},\ }\bibfield  {title} {\bibinfo {title} {Classification assessment
  methods},\ }\href {https://doi.org/10.1016/j.aci.2018.08.003} {\bibfield
  {journal} {\bibinfo  {journal} {Applied Computing and Informatics}\ }\textbf
  {\bibinfo {volume} {17}},\ \bibinfo {pages} {168} (\bibinfo {year}
  {2021})}\BibitemShut {NoStop}%
\bibitem [{\citenamefont {Grandini}\ \emph {et~al.}(2020)\citenamefont
  {Grandini}, \citenamefont {Bagli},\ and\ \citenamefont
  {Visani}}]{Grandini2020Metrics}%
  \BibitemOpen
  \bibfield  {author} {\bibinfo {author} {\bibfnamefont {M.}~\bibnamefont
  {Grandini}}, \bibinfo {author} {\bibfnamefont {E.}~\bibnamefont {Bagli}},\
  and\ \bibinfo {author} {\bibfnamefont {G.}~\bibnamefont {Visani}},\ }\href
  {https://doi.org/10.48550/arXiv.2008.05756} {\bibinfo {title} {Metrics for
  {{Multi-Class Classification}}: An {{Overview}}}} (\bibinfo {year} {2020}),\
  \Eprint {https://arxiv.org/abs/2008.05756} {arXiv:2008.05756 [stat]}
  \BibitemShut {NoStop}%
\bibitem [{\citenamefont {Birhanu}\ and\ \citenamefont
  {Jo}(2025{\natexlab{c}})}]{Birhanu2025Codesa}%
  \BibitemOpen
  \bibfield  {author} {\bibinfo {author} {\bibfnamefont {T.}~\bibnamefont
  {Birhanu}}\ and\ \bibinfo {author} {\bibfnamefont {H.-H.}\ \bibnamefont
  {Jo}},\ }\href@noop {} {\bibinfo {title} {Codes for the analysis framework
  for higher-order temporal correlations with applications to heartbeats}},\
  \bibinfo {howpublished} {https://github.com/tibebe22/hearbeat} (\bibinfo
  {year} {2025}{\natexlab{c}})\BibitemShut {NoStop}%
\end{thebibliography}

\end{document}